\documentclass[useAMS,usenatbib,twocolumn]{mn2e}
\usepackage{aas_macros}
\usepackage{amsmath,amssymb}
\usepackage{bm}
\usepackage{color}
\usepackage[dvipdfmx]{hyperref}
\usepackage{graphicx}
\usepackage{enumerate}
\usepackage{ulem}
\newcommand{\nnmb}{\nonumber\\}

\newcommand{\ds}{\displaystyle}
\newcommand{\mr}[1]{\mathrm{#1}}
\newcommand{\cmr}[1]{\,\mathrm{#1}}

\newcommand{\Hii}{H\,{\sc ii} }

\usepackage{amsmath}	% required for `\align' (yatex added)

\begin{document}
\author[K. Sugimura et al.
]{Kazuyuki Sugimura,$^1$\thanks{E-mail: sugimura@astr.tohoku.ac.jp} 
Takashi Hosokawa,$^{2}$
Hidenobu Yajima,$^{1,3}$ \newauthor
Kohei Inayoshi$^{4}$ and Kazuyuki Omukai$^{1}$
\\
$^1$Astronomical Institute, Tohoku University, Aoba, Sendai 980-8578,
Japan \\ $^2$Department of Physics, Kyoto University, Sakyo, Kyoto
606-8502, Japan\\ $^3$Frontier Research Institute for Interdisciplinary
Sciences, Tohoku University, Aoba, Sendai 980-8578, Japan\\
$^4$Department of Astronomy, Columbia University, 550 W. 120th Street,
New York, NY 10027, USA } 
\title[The effect of angular momentum and radiation on accretion] {Stunted accretion growth of black holes by combined effect of the flow angular
momentum and radiation feedback} \maketitle
\topmargin-1cm

\begin{abstract}
Accretion on to seed black holes (BHs) is believed to play a crucial
role in formation of supermassive BHs observed at high-redshift ($z>6$).
Here, we investigate the combined effect of gas angular momentum and
radiation feedback on the accretion flow, by performing 2D axially
symmetric radiation hydrodynamics simulations that solve the flow
structure across the Bondi radius and the outer part of the accretion
disc simultaneously. The accreting gas with finite angular momentum
forms a rotationally-supported disc inside the Bondi radius, where the
accretion proceeds by the angular momentum transport due to assumed
$\alpha$-type viscosity.  We find that the interplay of radiation and
angular momentum significantly suppresses accretion even if the
radiative feedback is weakened in an equatorial shadowing region.  The
accretion rate is $O(\alpha)\sim O(0.01\,\text{--}\,0.1)$ times the Bondi value,
where $\alpha$ is the viscosity parameter.  By developing an analytical
model, we show that such a great reduction of the accretion rate
persists unless the angular momentum is so small that the corresponding
centrifugal radius is $\lesssim 0.04$ times the Bondi radius.  We argue that
BHs are hard to grow quickly via rapid mass accretion considering the
angular momentum barrier presented in this paper.

\end{abstract}

%%%%%%%%%%%%%%%%%%%%%%%%%%%%%%%%%%%%%%%%%%%%%%%%%%%%%
%%%%%%%%%%%%%%%%%%%%%%%%%%%%%%%%%%%%%%%%%%%%%%%%%%%%%
\section{Introduction}
\label{sec:intro}
%%%%%%%%%%%%%%%%%%%%%%%%%%%%%%%%%%%%%%%%%%%%%%%%%%%%%
%%%%%%%%%%%%%%%%%%%%%%%%%%%%%%%%%%%%%%%%%%%%%%%%%%%%%

Observations of supermassive black holes (SMBHs) with mass $\sim
10^9\,M_\odot$ at redshift $z\gtrsim6$, or $\lesssim 1\cmr{Gyr}$ after
the big bang, severely constrain their formation mechanism (e.g.,
\citealp{Fan:2001aa,Willott:2010aa,Mortlock:2011aa,Venemans:2013aa,Wu:2015aa};
see also \citealp{Gallerani:2017aa} for review).  Heavy seed BHs have
been invoked in several scenarios to reconcile the short available time
with their hugeness \citep[see, e.g.,][for a
review]{Volonteri:2012ab,Haiman:2013aa}, which includes (1) Pop III
remnant BHs with mass of $M_\mr{BH}\lesssim 10^3\,M_\odot$
\citep[e.g.,][]{Yoshida:2008aa,
Hosokawa:2011aa,Hosokawa:2016aa,Susa:2014aa,Hirano:2015aa,Stacy:2016aa};
(2) direct collapse BHs with $M_\mr{BH}\sim 10^5\,M_\odot$ formed via
the collapse of supermassive stars
\citep[e.g.,][]{Omukai:2001aa,Bromm:2003aa,Hosokawa:2012aa,
Sugimura:2014aa,Sugimura:2016aa,Inayoshi:2014aa,Chon:2016aa,Umeda:2016aa};
and (3) massive BHs with $M_\mr{BH}\sim 10^3\,M_\odot$ formed as a
consequence of stellar mergers in dense clusters
\citep[e.g.,][]{Omukai:2008aa,Devecchi:2009aa,Katz:2015ab,Tagawa:2015aa,Yajima:2016aa,Sakurai:2017aa}.

In all the cases, the seeds have to increase their mass further by
several orders of magnitude.  Although Pop III remnants with relatively
small mass have been claimed to become SMBHs by $z\gtrsim6$ by very
rapid (super-Eddington) gas accretion
\citep{Volonteri:2005aa,Madau:2014aa,Alexander:2014aa,Volonteri:2015aa}, whether this
actually occurs or not is still very uncertain.  This is because our
knowledge on the realistic accretion process is still limited.

\cite{Bondi:1952aa} analytically investigated the most simplistic case
of steady, spherical accretion of the polytropic gas from a homogeneous
and static medium \citep[see also, e.g.,][]{Shapiro:1983aa}.  For a BH
with mass $M_\mr{BH}$ surrounded by a gas with density $\rho_\infty$ and
sound speed $c_\mr{s,\infty}$, the so-called Bondi accretion rate is
\begin{align}
&\dot{M}_\mr{B}
=\frac{4\pi\lambda_\mr{B}\rho_\infty G^2 M_\mr{BH}^2}{c_\mr{s,\infty}^3}
=1.7\times 10^{-3}\left(\frac{\lambda_\mr{B}}{1.12}\right)
\nnmb
\times\,&\left(\frac{n_\mr{H,\infty}}{10^5\cmr{cm^{-3}}}\right)
 \left(\frac{M_\mr{BH}}{10^3\,M_\odot}\right)^{2}
\left(\frac{c_\mr{s,\infty}}{8\cmr{km\, s^{-1}}}\right)^{-3}
M_\odot\cmr{yr^{-1}}
 \,.
\label{eq:6}
\end{align}
Here, $\lambda_\mr{B}$ is the non-dimensional factor depending on the
polytropic index $\gamma$ ($\lambda_\mr{B}=1.12$ for the isothermal,
i.e., $\gamma=1$, case) and $c_\mr{s,\infty}= 8\cmr{km\, s^{-1}}$ for a
neutral primordial gas with temperature $T=10^4\cmr{K}$.  In the second
expression of Eq.~\eqref{eq:6}, we use the relation between the number
density of hydrogen atoms $n_\mr{H,\infty}$ and $\rho_\infty$ for the
primordial gas.

Although the Bondi rate provides a rough estimate of an accretion rate
onto a BH, various effects reduce the rate in realistic situations. For
instance, radiation feedback from circum-BH discs can significantly
disturb the accretion flow so that the rate decreases down to $\lesssim
0.01\dot{M}_\mr{B}$
\citep[e.g.,][]{Milosavljevic:2009ab,Park:2011aa,Park:2012aa,Park:2017ab}. Recent
studies have proposed pathways to alleviate the feedback, in such cases
as where the radiation is highly obscured around the equatorial plane by
disc winds \citep{Sugimura:2017ab,Takeo:2018aa} or where the radiation
is trapped in a dense (super-Eddington) accretion flow
\citep{Inayoshi:2016ac,Sakurai:2016aa}. These studies, however, do not
consider the effect of angular momentum on the large-scale accretion
flow, assuming that the size of the circum-BH disc is much smaller than
the Bondi radius.  This assumption may not be always the case. We need
to know the condition that the effect of angular momentum becomes
significant.

Accretion to active galactic nuclei (AGNs) has been investigated in the
last few decades
\citep[e.g.,][]{Ciotti:2001aa,Wada:2002aa,Kawakatu:2008aa,Kurosawa:2009aa,Novak:2011aa,Kawaguchi:2010aa,Barai:2012aa,
Yuan:2012aa}. In such a context, the effect of gas angular momentum has partly
been investigated.  Even without radiation feedback, moderate gas
angular momentum significantly suppresses the accretion in
low-luminosity AGN \citep[][]{Proga:2003aa,
Proga:2003ab,Cuadra:2006aa,Li:2013aa,Gaspari:2015aa,Inayoshi:2018aa}.
This implies that the effect of angular momentum, as well as radiation
feedback, needs to be considered in studying the accretion on to the
seed BHs. In \cite{Begelman:2017aa}, they have studied the
properties of super-Eddington accretion flows for cases where the disc
size is smaller than the photon-trapping radius. In contrast, we are interested in cases where the disc is much larger than the
photon-trapping radius, which is far smaller than the Bondi radius.

In this work, to see the combined effect of the angular momentum and the
radiation feedback on seed BH accretion, we perform a set of 2D
axisymmetric radiation hydrodynamics simulations, considering both
finite gas angular momentum and radiation from the circum-BH discs.  We
follow formation of rotationally-supported discs and subsequent viscous
accretion through them. We find that angular momentum of the gas, in
cooperation with radiation feedback, suppresses the accretion on to seed
BHs. To understand its mechanism, we also develop an analytical model
for accretion from a rotating medium.

The paper is organized as follows. In Sec.~\ref{sec:num_setting}, we
describe the numerical method and the parameter sets explored. In
Sec.~\ref{sec:results}, we present the results of our simulations. In
Sec.~\ref{sec:formulation}, we develop an analytical model, which allows
us to obtain the condition for suppression of accretion.  In
Sec.~\ref{sec:discussion}, we discuss the possible growth history of Pop
III remnant BHs based on our findings, as well as the caveats of our
simulations.  Finally, we summarize in Sec.~\ref{sec:conclusion}.

%%%%%%%%%%%%%%%%%%%%%%%%%%%%%%%%%%%%%%%%%%%%%%%%%%%%%
%%%%%%%%%%%%%%%%%%%%%%%%%%%%%%%%%%%%%%%%%%%%%%%%%%%%%
\section{Method}
\label{sec:num_setting}
%%%%%%%%%%%%%%%%%%%%%%%%%%%%%%%%%%%%%%%%%%%%%%%%%%%%%
%%%%%%%%%%%%%%%%%%%%%%%%%%%%%%%%%%%%%%%%%%%%%%%%%%%%%

We perform axisymmetric 2D radiation hydrodynamics (RHD) simulations, by
using a modified version of a public grid-based multidimensional
magnetohydrodynamics (MHD) code {\tt PLUTO 4.1} \citep{Mignone:2007aa},
which is mostly the same as used in our previous work
(\citealp{Sugimura:2017ab}; see also
\citealp{Kuiper:2010aa,Kuiper:2010ab,Kuiper:2011aa,Kuiper:2013aa,Hosokawa:2016aa}).
Major update here is implementation of physics related to the rotation
of gas.  In the rest of this paper, we use both spherical $(r,\ \theta,\
\phi)$ and cylindrical $(R,\ z,\ \phi)$ coordinates interchangeably,
although the spherical one was actually used in the calculation.

%%%%%%%%%%%%%%%%%%%%%%%%%%%%%%%%%%%%%%%%%%%%%%%%%%%%%
\subsection{Modelling accretion on to a BH under radiation feedback}
\label{sec:rad_fb}
%%%%%%%%%%%%%%%%%%%%%%%%%%%%%%%%%%%%%%%%%%%%%%%%%%%%%

\begin{figure}
\centering \includegraphics[width=6cm]{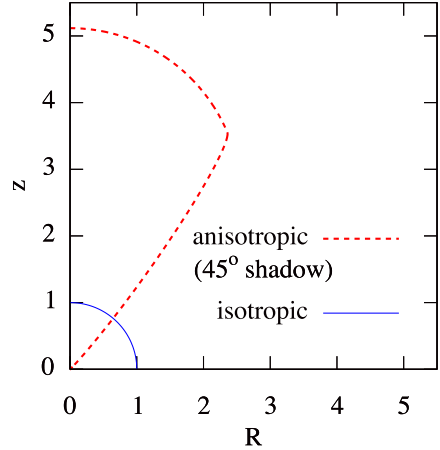}
 \caption{Directional dependence of ionizing flux.  The radial extent
 represents the strength of the flux compared to the isotropic case.}  \label{fig:dirdep}
\end{figure}

We briefly describe the basic properties of the code that are common to our previous work \citep[see][for details]{Sugimura:2017ab}. 
With this code, we follow BH accretion from a homogeneous surrounding
medium under radiation feedback.

We put a BH at the centre of computational domain and treat it as a
sink. Through the sink surface, the gas is allowed to flow in, while the
radiation is emitted according to a semi-analytical model described
below.  We solve the coupled equations of hydrodynamics, radial
multi-frequency radiation transport and primordial gas chemistry.  In
the current version of the code, helium ionization is considered,
hydrogen molecules are assumed to be completely destroyed, and the
self-gravity of gas is ignored.

We consider ionizing photons emitted from the circum-BH discs with a
semi-analytical prescription. Using the accretion rate $\dot{M}$
evaluated at the sink surface, the luminosity $L$ is given by
\citep{Watarai:2000aa}
\begin{align}
 L = 
\begin{cases}
 2\, L_\mr{E}\,\left[1+\ln\left(\dot{m}/20\right)\right] & \dot{m} > 20\\
0.1\, L_\mr{E}\, \dot{m}  & \dot{m} < 20
\end{cases}\,,
\label{eq:3}
\end{align}
with $\dot{m}\equiv \dot{M}/\dot{M}_\mr{E}$.  Here, $L_\mr{E}$ is the
Eddington luminosity,
\begin{align}
L_\mr{E}=\frac{4\pi G M_\mr{BH} c m_\mr{p}}{\sigma_\mr{T}}
&=3.3\times 10^7
\left(\frac{M_\mr{BH}}{10^3M_\odot}\right) L_\odot\,,
\label{eq:7}
\end{align}
and $\dot{M}_\mr{E}$ the (efficiency-independent) Eddington accretion
rate,
\begin{align}
 \dot{M}_\mr{E}
&=\frac{L_\mr{E}}{c^2}
=2.2\times 10^{-6} \left(\frac{M_\mr{BH}}{10^3M_\odot}\right)\,M_\odot\, \mr{yr^{-1}}\,.
\label{eq:10}
\end{align}
 Note that the decrease of radiative efficiency at $\dot{M}\gg
\dot{M}_\mr{E}$ is caused by photon trapping in the slim discs
\citep[][]{Abramowicz:1988aa}.  We assume that the spectral energy
distribution is given by the power law $L_\nu\propto \nu^{-1.5}$ with
the frequency minimum at $h\nu_\mr{min}=13.6\cmr{eV}$ and that
$L=\int_{\nu_\mr{min}}^{\infty}L_\nu\,\mathrm{d}\nu$
\citep[e.g.,][]{Park:2011aa}.

Radiation from the sink is supposed to be anisotropic, partly because
the photons from the hot inner part of the circm-BH disc is obscured by
some outer materials.  For example, \cite{Proga:2000aa} suggested that
line-driven winds from an SMBH accretion disc make a shadowing region
with opening angle from the equatorial plane $\sim 10^\circ$.  However,
especially in the case with smaller BH mass or lower metallicity, the
anisotropy is highly uncertain due to the lack of knowledge on obscuring
materials, which are presumably (failed) winds or coronae above the
disc\footnote{Note that the anisotropy generated in this way exists
independently of the self-shielding effect in slim discs.}  \citep[see,
e.g.,][]{Begelman:1983aa,Hollenbach:1994aa,Woods:1996aa,Proga:2000aa,Wada:2012aa,Suzuki:2014aa,Nomura:2016aa}.
We thus do not attempt to realistically model the anisotropy, but
instead, consider the two limiting cases: isotropic radiation and
anisotropic radiation with a $\sim 45^\circ$ shadowing region
\citep[Fig.~\ref{fig:dirdep}; Equation~9 of][]{Sugimura:2017ab}.  We
expect the reality lies somewhere between the two.

%%%%%%%%%%%%%%%%%%%%%%%%%%%%%%%%%%%%%%%%%%%%%%%%%%%%%
\subsection{Modelling gas rotation}
\label{sec:ang_mom}
%%%%%%%%%%%%%%%%%%%%%%%%%%%%%%%%%%%%%%%%%%%%%%%%%%%%%

We assume that the surrounding gas has the initial profile of specific
angular momentum
\begin{align}
 j(\bf{x}) = 
\begin{cases}
\ds \left(\frac{R}{r_\mr{B}}\right)^2j_\infty
& R < r_\mr{B} 
\\
j_\infty		       
& r_\mr{B} < R 
\end{cases}\,,
\label{eq:29}
\end{align}
with the Bondi radius
\begin{align}
\!\! r_\mr{B}&=\frac{G M_\mr{BH}}{c_\mr{s,\infty}^2}
=1.4\times10^4\bigg(\!\frac{M_\mr{BH}}{10^3\,M_\odot}\!\bigg)\!
\bigg(\!\frac{c_\mr{s,\infty}}{8\cmr{km\, s^{-1}}}\!\bigg)^{\!-2}\!\!\cmr{au}\,.
\label{eq:15}
\end{align}
That is, the gas has constant $j$ throughout the computational domain,
except near the rotation axis with $R < r_\mr{B}$, where the angular
velocity $\Omega=j/R^2$ is constant instead.  Below, we interchangeably
use $j$ and the centrifugal radius $R_\mr{c}$, where the centrifugal
force and the BH gravity balances, to indicate the angular momentum of
the accreted gas, as $R_\mr{c}$ is related to $j$ as
\begin{align}
 R_\mr{c}=\frac{j^2}{GM_\mr{BH}}\,.  \label{eq:11}
\end{align}

Angular momentum must be transported for a gas with finite $j$ to reach
the vicinity of the BH.  We assume the $\alpha$-type viscous stress
(\citealp{Shakura:1973aa}; see also
\citealp{Igumenshchev:1999aa,Stone:1999aa,Li:2013aa}), which is possibly
due to turbulence driven by the magnetorotational instability
\citep[MRI;][]{Balbus:1998aa}.  The gravitational torque is
insignificant in the cases studied here, because the disc is highly
gravitationally stable (see Appendix~\ref{sec:toomreQ_iso}).  Note that
while the value of $\alpha$ corresponding to the actual MRI turbulence
is yet to be fully understood, \cite{Bai:2013aa}, for example, have
reported that $\alpha$ is $\sim 0.01\text{--}0.02$ in the case with a
weak vertical magnetic field but can be even larger than unity in the
strong field case.

To assure that the viscosity works only inside the disc
\citep[e.g.,][]{Stone:1999aa}, we introduce a confinement factor $f$ and
use the following expression for
viscosity:\footnote{Equation~\eqref{eq:16} can also be written as
$\nu=f\,\alpha \,c_\mr{s,ad}^2/\Omega_\mr{K}$, with adiabatic sound
speed $c_\mr{s,ad} = \sqrt{\gamma}\,c_\mr{s}$.}
 \begin{align}
 \nu = f\left(\frac{\Omega}{\Omega_\mr{K}}\right)\, \frac{\alpha\,\gamma\,c_\mr{s}^2}{\Omega_\mr{K}}\,,
\label{eq:16}
\end{align}
with isothermal sound speed $c_\mr{s}$, adiabatic index $\gamma=5/3$ and
Keplerian angular velocity $\Omega_\mr{K} = \sqrt{GM_\mr{BH}/R^3}$.
Here, we identify the disc region based on the degree of rotational
support against the BH gravity: we set $f=0$ for
$\Omega/\Omega_\mr{K}<\tilde{\Omega}_\mr{th}$ and $f=1$ for
$\tilde{\Omega}_\mr{th}+\Delta\tilde{\Omega}<\Omega/\Omega_\mr{K}$, and
linearly interpolate between them, with the threshold value
$\tilde{\Omega}_\mr{th}=0.8$ and the transition width
$\Delta\tilde{\Omega}=0.1$ in most of our calculations. We check the
effect of changing $\tilde{\Omega}_\mr{th}$ in
Sec.~\ref{sec:std_acc_disc}. Note that we have seen in test runs that if
the viscosity is not limited to the disc, the disc will disappear and
the flow becomes spherical due to angular momentum transport in the
entire computational domain.

%%%%%%%%%%%%%%%%%%%%%%%%%%%%%%%%%%%%%%%%%%%%%%%%%%%%%
\subsection{Cases considered}
\label{sec:cases}
%%%%%%%%%%%%%%%%%%%%%%%%%%%%%%%%%%%%%%%%%%%%%%%%%%%%%

\begin{table}
 \centering
 \caption{Summary of runs.}
 \label{tab:model}
 \begin{tabular}{ccccc} \hline
$R_\mr{c,\infty}/r_\mr{B}$& $\alpha$ &
radiation& 
$t_\mr{end}\,[\mr{yr}]$&
$\dot{M}/\dot{M}_\mr{B}$$^{c,d}$
 \\\hline
0.1 & 0.01 & anisotropic$^{a}$ &  $1.2\times10^6$&$2.9\times10^{-3}$ \\
0.1 & 0.1 & anisotropic$^{a}$ &  $1.2\times10^6$&$5.2\times10^{-2}$ \\
0.3 & 0.01 & anisotropic$^{a}$ & $1.2\times10^6$&$1.2\times10^{-3}$ \\
0.03 & 0.01 & anisotropic$^{a}$ & $1.2\times10^6$&$4.3\times10^{-1}$ \\

0.1 & 0.01 & isotropic & $4\times10^5$&$1.1\times10^{-4}$ \\
0.1 & 0.1 & isotropic & $4\times10^5$&$2.6\times10^{-4}$ \\
0.3 & 0.01 & isotropic & $4\times10^5$&$9.7\times10^{-5}$ \\

0.1 & 0.01 & no & $2\times10^5$&$4.1\times10^{-2}$ \\
0.1 & 0.1 & no & $2\times10^5$&$1.2\times10^{-1}$ \\
0.3 & 0.01 & no & $2\times10^5$&$3.4\times10^{-2}$ \\
0.03 & 0.01 & no & $2\times10^5$&$8.1\times10^{-1}$ \\\hline

\multicolumn{5}{c}{Non-rotating case$^{b}$}\\ \hline
0 & --- & anisotropic$^{a}$ &  $2\times10^6$&$5.9\times10^{-1}$ \\
0 & --- & isotropic & $5\times10^5$&$1.7\times 10^{-3}$ \\
0 & --- & no &--- &$1$ \\\hline
 \end{tabular}\\
 \begin{flushleft}
NOTES.\textemdash We set $M_\mr{BH}=10^3\,M_\odot$,
$n_\mr{H,\infty}=10^5\cmr{cm^{-3}}$ and $T_\mr{\infty}=10^4\cmr{K}$ in
all runs.\\
$^{a}$ Anisotropic radiation with $\sim 45^\circ$ shadow. \\ 
$^{b}$ Results from analytical estimate and the simulations in
\cite{Sugimura:2017ab}.\\
$^{c}$ Accretion rate at the end of simulation (averaged near the end of
simulation if oscillating).\\ 
$^{d}$ The accretion rate can also be normalized by $\dot{M}_\mr{E}$
using the relation $\dot{M}/\dot{M}_\mr{E}\approx
800\dot{M}/\dot{M}_\mr{B}$ (See Eqs.~\ref{eq:6} and
\ref{eq:10}).
 \end{flushleft}
\end{table}

Our runs are summarized in Table~\ref{tab:model}.  In all runs, we set
$M_\mr{BH}=10^3\,M_\odot$ and fix it constant during the calculations.
The surrounding medium is assumed to be the neutral primordial gas with
density $n_\mr{H,\infty}=10^5\cmr{cm^{-3}}$ and temperature
$T_\infty=10^4\cmr{K}$. In the fiducial run, we assume anisotropic
radiation field with $\alpha=0.01$ and $R_\mr{c,\infty}\equiv
j_\infty^2/GM_\mr{BH} =0.1\,r_\mr{B}$.  In order to examine the
parameter dependence, we study cases with different parameters of
$(\alpha,\ R_\mr{c,\infty}/r_\mr{B})=(0.01,\ 0.03)$, $(0.1,\ 0.1)$ and
$(0.01,\ 0.3)$. For comparison, we also solve the flow structure under
the isotropic radiation field with the similar parameter sets.

We start from the homogeneous and quasi-static ($v_r=v_\theta=0$)
initial condition. We turn the radiation off for the first
$2\times10^5\cmr{yr}$ to allow the flow to settle in a steady state with
a rotationally-supported disc.  Note that this period is longer than
either the dynamical time at $r_\mr{B}$,
$t_\mr{B}=\Omega_\mr{K}^{-1}(r_\mr{B})\sim 10^{4}\cmr{yr}$, or the
viscous time at $R_\mr{c,\infty}$, $t_\mr{visc,c}=
R_\mr{c,\infty}^2/\nu(R_\mr{c,\infty}) \sim
10^{5}(\alpha/0.01)^{-1}(R_\mr{c,\infty}/0.1 r_\mr{B})^{1/2}\cmr{yr}$.
We then turn on the radiation and follow the evolution until
$t_\mr{end}=1.2\times10^6\cmr{yr}$ in the anisotropic radiation runs and
$4\times10^5\cmr{yr}$ in the isotropic radiation runs.

We impose the reflection symmetry with respect to the equatorial plane,
as well as axisymmetry around the rotation axis, and thus $\theta$
extends from $0^\circ$ to $90^\circ$. In the $r$-direction, our
computational domain ranges from $r_\mr{in}=10^{-2}\,r_\mr{B}$ to
$r_\mr{out}=10^{2}\,r_\mr{B}$.  Note that $r_\mr{in}$ is smaller than
either $R_\mr{c,\infty} \sim 10^{-1}\,r_\mr{B}$ (see above) or the Bondi
radius for \Hii gas $r_\mr{B,HII} \sim10^{-1}\,r_\mr{B}$ with $\sim
7\times10^4\cmr{K}$, while $r_\mr{out}$ is larger than the size of \Hii
region.  We use logarithmic grids in the $r$ direction and homogeneous
ones in the $\theta$ directions, with $N_r\times N_\theta = 512\times
144$.  We discuss the dependence of our results on the numerical
configuration in Appendix~\ref{sec:resdep_iso}.

At the outer boundary, we let the flow go out from the computational
domain but not come into it. At the inner boundary, we impose the same
boundary condition in most cases, but when the gas is judged to belong
to the Keplerian accretion disc (specifically, when $v_{r,\mr{in}}<0$
and $\Omega_\mr{in}>0.95\,\Omega_\mr{K,in}$, where quantities with
subscripts ``in'' are evaluated at $r_\mr{in}$), we determine the
physical quantities in the ghost cells according to the radial profile
of the isothermal Keplerian disc: $\rho=
(r/r_\mr{in})^{-3}\rho_\mr{in}$, $p= (r/r_\mr{in})^{-3}p_\mr{in}$,
$v_r=(r/r_\mr{in})^{1/2}v_{r,\mr{in}}$,
$v_\phi=(r/r_\mr{in})^{-1/2}v_{\phi,\mr{in}}$ and
$\mathcal{Q}=\mathcal{Q}_\mr{in}$ for the other variables (see
Appendix~\ref{sec:std_acc_disc}).
We set a temperature floor at $T_\mr{min}=10^4\cmr{K}$ for simplicity
\citep[e.g.,][]{Sugimura:2017ab}, as well as minimum density
$n_\mr{min}=10^{-1}\cmr{cm^{-3}}$ and maximum velocity
$v_\mr{max}=150\cmr{km/s}$ for numerical reasons.  For the lowest angular momentum ($R_\mr{c,\infty}/r_\mr{B}=0.03$) run, we additionally limit $j$ to below $j_\infty$, because otherwise the disc accretes the gas with $j>j_\infty$, receiving additional angular momentum transported from the inner part of the disc. Recall that our aim here is to study how the accretion flow structure varies with different $j_\infty$ in a well controlled manner. We discuss how such a situation may be realized in Sec.~\ref{sec:caveats} later. For the other higher angular momentum ($R_\mr{c,\infty}/r_\mr{B}\geq 0.1$) cases, we do not put the same upper limit on $j$ because it hardly changes the result.

\begin{figure}
\centering \includegraphics[width=8cm]{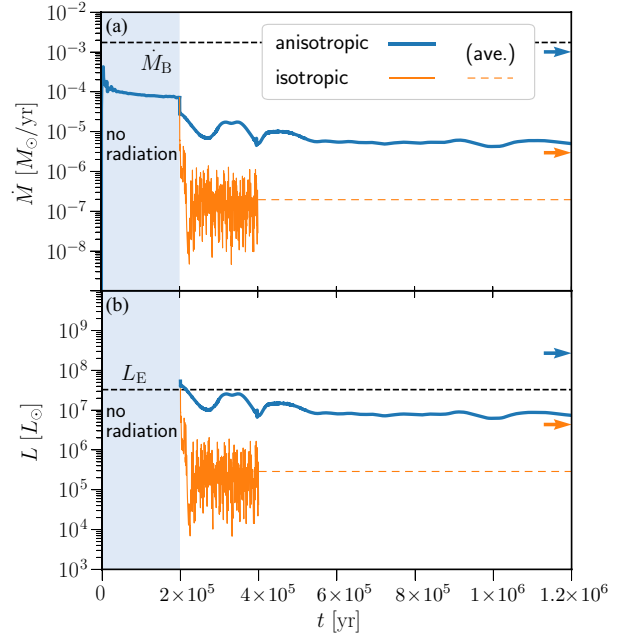}
\caption{Time evolution of the (a) accretion rate and (b) luminosity in
the runs with $\alpha=0.01$ and $R_\mr{c,\infty}=0.1\,r_\mr{B}$.  Both
fiducial anisotropic radiation case (blue) and isotropic radiation case
(orange) are shown.  In these runs, the radiation is turned off for the
initial $2\times 10^5$ years.  The bottom thin-dashed lines represent
the averaged values for $3\times 10^5\cmr{yr}<t<4\times 10^5\cmr{yr}$.
The results for the non-rotating case \citep{Sugimura:2017ab} are also
shown by arrows for comparison.  } \label{fig:mdot_L_FB_fid}
\end{figure}

%%%%%%%%%%%%%%%%%%%%%%%%%%%%%%%%%%%%%%%%%%%%%%%%%%%%%
%%%%%%%%%%%%%%%%%%%%%%%%%%%%%%%%%%%%%%%%%%%%%%%%%%%%%
\section{Results}
\label{sec:results}
%%%%%%%%%%%%%%%%%%%%%%%%%%%%%%%%%%%%%%%%%%%%%%%%%%%%%
%%%%%%%%%%%%%%%%%%%%%%%%%%%%%%%%%%%%%%%%%%%%%%%%%%%%%

Results of our simulations are summarized in Table~\ref{tab:model},
along with our previous cases for the non-rotating gas
\citep{Sugimura:2017ab}, for comparison. Below, we first describe the
fiducial anisotropic radiation run in Sec.~\ref{sec:fiducial}. We then
present the isotropic radiation run with the same parameter set in
Sec.~\ref{sec:rad_iso} and finally see the cases with different
parameter sets in Sec.~\ref{sec:iso_parm}.

%%%%%%%%%%%%%%%%%%%%%%%%%%%%%%%%%%%%%%%%%%%%%%%%%%%%%
\subsection{The fiducial run}
\label{sec:fiducial}
%%%%%%%%%%%%%%%%%%%%%%%%%%%%%%%%%%%%%%%%%%%%%%%%%%%%%

\begin{figure}
\centering \includegraphics[width=6cm]{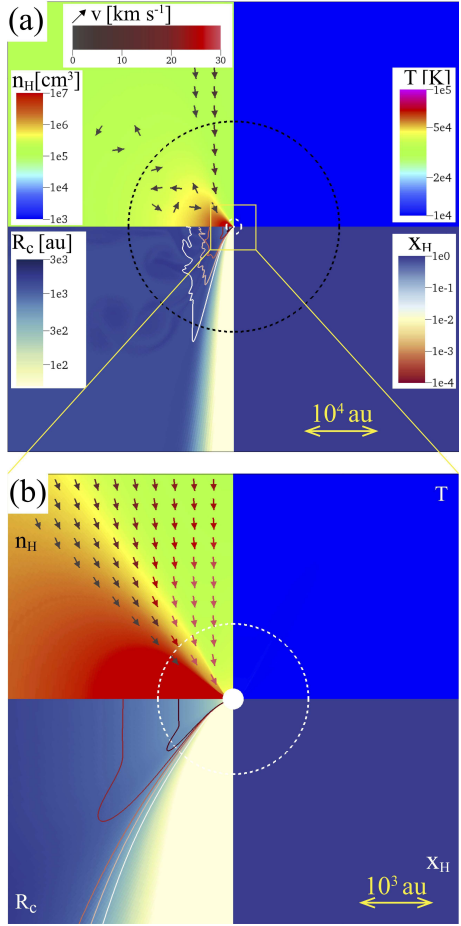}
\caption{ The gas distribution on the scales of (a) $10^4\cmr{au}$ and
(b) $10^3\cmr{au}$ just before turning on the radiation in the fiducial
anisotropic radiation run.  In each panel, the four quadrants (clockwise
from top left) represent number density $n_\mr{H} \cmr{[cm^3]}$,
temperature $T \cmr{[K]}$, neutral fraction of hydrogen $x_\mr{H}$ and
specific angular momentum shown by the corresponding centrifugal radius
$R_\mr{c}\,(=j^2/GM_\mr{BH})\cmr{[au]}$. The arrows represent the
velocity vector $\bf{v}$, shown only when $|\bf{v}| > 1\cmr{km
s^{-1}}$. The contours in the bottom left panel represent
$\Omega/\Omega_\mr{K}=$ 0.5 (white), 0.6 (pink), 0.7 (orange), 0.8 (red)
and 0.9 (dark red). The Bondi radii for neutral and ionized gases are
shown as dashed black and white circles, respectively.}
\label{fig:NRc01a1em2}
\end{figure}
\begin{figure}
\centering \includegraphics[width=6cm]{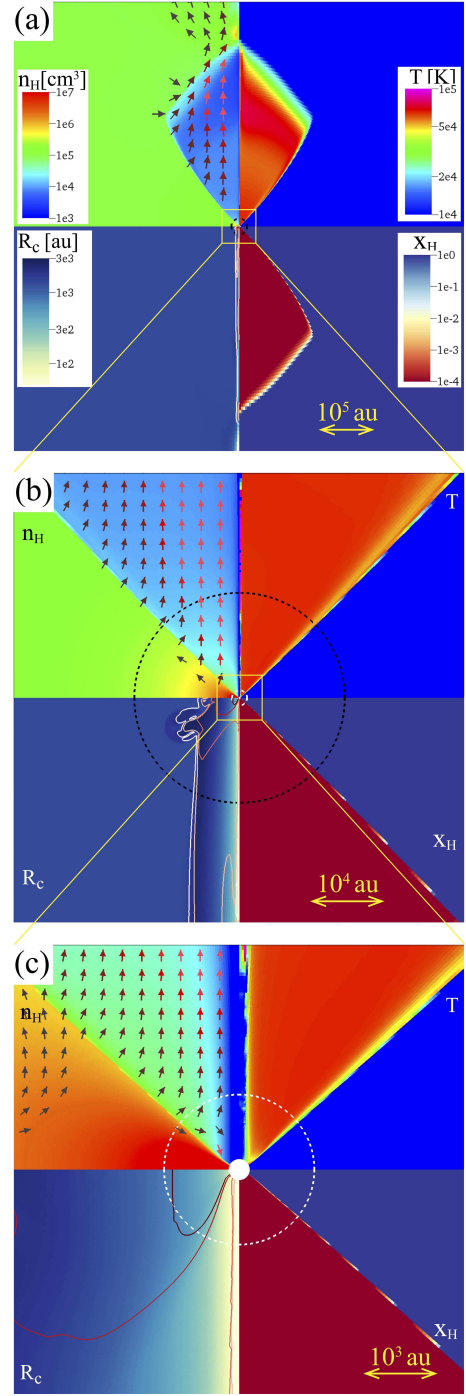}
\caption{Same as Fig.~\ref{fig:NRc01a1em2} but at the end of the
simulation, when the gas in the polar regions are photoionized by the
anisotropic radiation. Here, the gas distribution is plotted on the
scales of (a) $10^5\cmr{au}$, (b) $10^4\cmr{au}$ and (c) $10^3\cmr{au}$.
} \label{fig:RSc01a1em2}
\end{figure}

We first describe the fiducial run, where the parameter set is given as
follows: $M_\mr{BH}=10^3\,M_\odot$, $n_\mr{H,\infty}=10^5\cmr{cm^{-3}}$,
$T_\infty=10^4\cmr{K}$, $\alpha=0.01$ and
$R_\mr{c,\infty}=0.1\,r_\mr{B}$.  The radiation field is assumed to be
anisotropic with $\sim 45^\circ$ shadow (see
Fig.~\ref{fig:dirdep}). Such a wide obscuration can be regarded as an
extreme of strong anisotropy.  Starting from the homogeneous initial
condition, we follow the evolution of flow with the radiation turned off
until $2\times 10^5\cmr{yr}$.  We then turn on the radiation and follow
the evolution until $t_\mr{end}=1.2\times10^6\cmr{yr}$.

Figs.~\ref{fig:mdot_L_FB_fid}(a) and (b) show the time evolution of the
accretion rate $\dot{M}$ and the luminosity $L$, respectively, along
with the result for the case with isotropic radiation, which will be
described in the next section.  Fig.~\ref{fig:mdot_L_FB_fid}(a) shows
that during the early period without radiation the accretion rate
$\dot{M}$ once converges to the constant value
$7.2\times10^{-5}M_\odot\cmr{yr^{-1}}$, only less than $1/10$ of the
Bondi rate $\dot{M}_\mr{B}$ ($=1.7\times 10^{-3}M_\odot\cmr{yr^{-1}}$).
This remarkable reduction in $\dot{M}$ is totally attributable to the
effect of angular momentum.

\begin{figure*}
\centering \includegraphics[width=6cm]{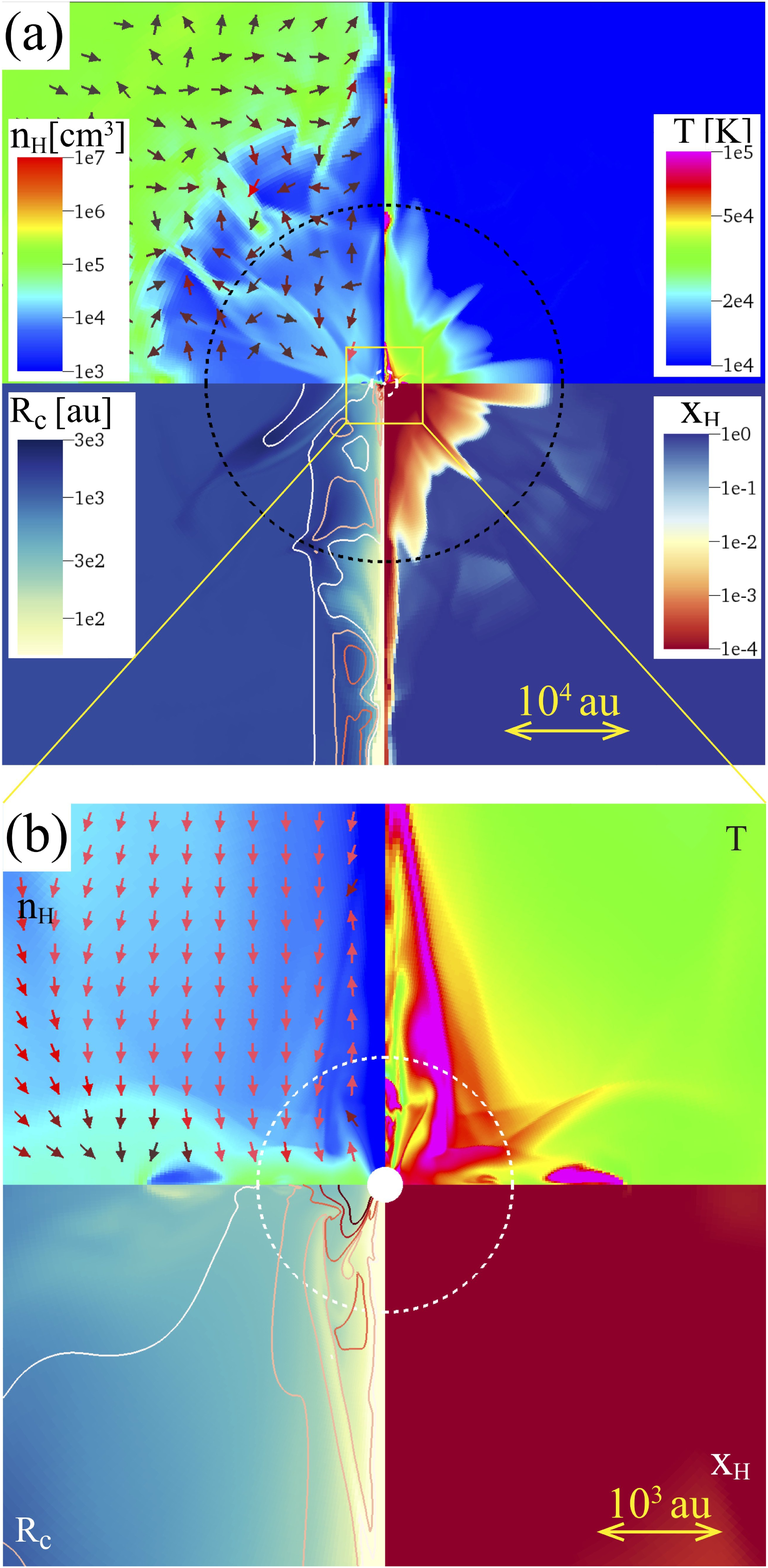}
\hspace{0.7cm} \centering
\includegraphics[width=6cm]{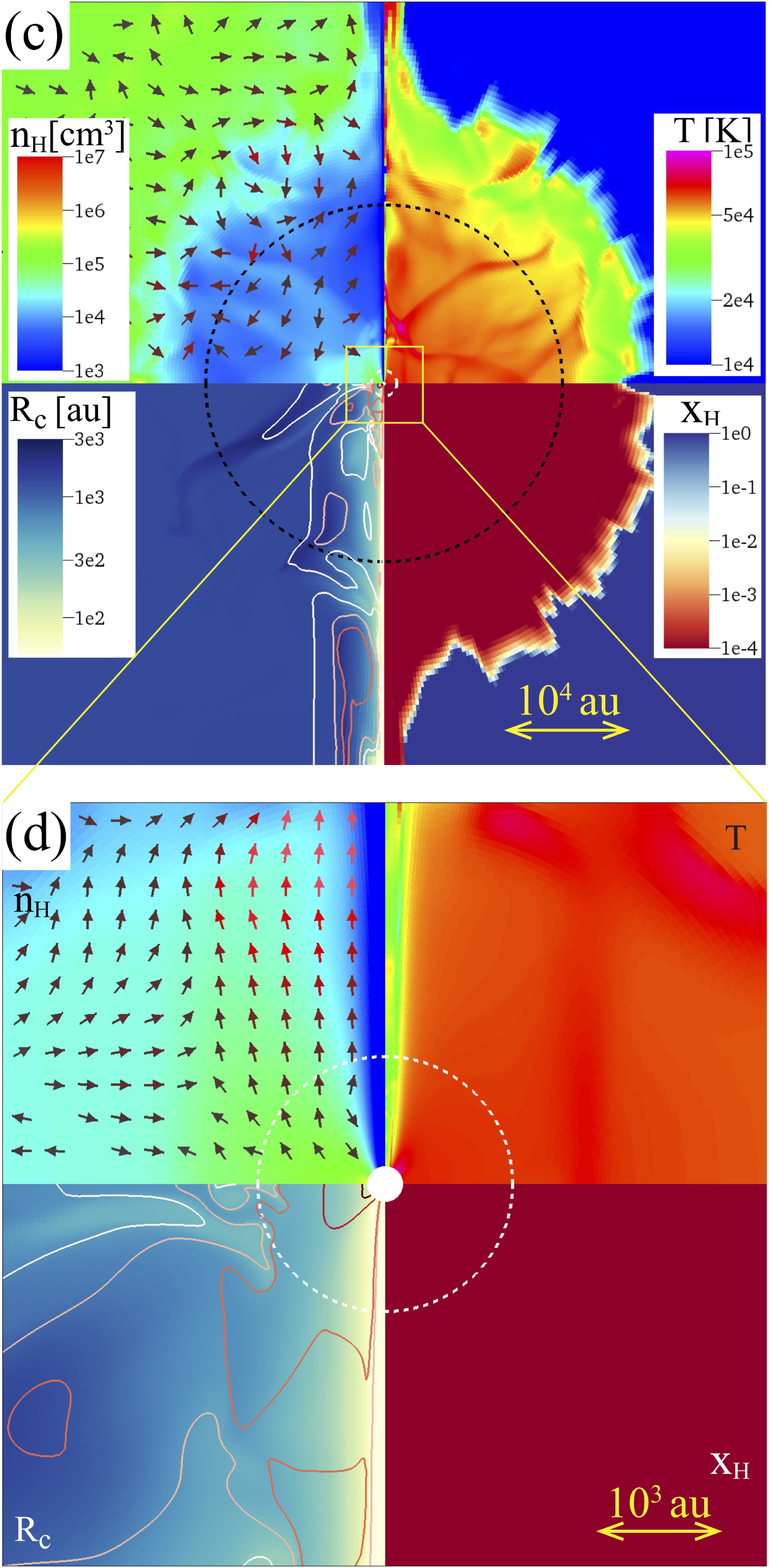} \caption{Same
as Fig.~\ref{fig:NRc01a1em2} but (a, b) before and (c, d) after an
accretion burst in the isotropic radiation run. }
\label{fig:RIc01a1em2_BBAB}
\end{figure*}

Fig.~\ref{fig:NRc01a1em2} shows the gas distribution just before turning
on the radiation.  A flared disc formed inside $r_\mr{B}$ can be
seen. In the disc, the gravity is balanced with the centrifugal force
inside $R_\mr{c,\infty}$ and with the pressure gradient outside
$R_\mr{c,\infty}$, so that the dynamical equilibrium ($v_r=v_\theta=0$)
is approximately maintained throughout.  In the bipolar regions, a gas
with low angular momentum directly flows into the sink without hitting
on the disc.

Before turning on the radiation, the gas is almost isothermal at $\sim
10^4\cmr{K}$ due to the efficient Ly$\alpha$ cooling throughout the
computational domain (see Fig.~\ref{fig:NRc01a1em2}).  As an experiment,
we have rerun the isothermal simulation setting $T=10^4\cmr{K}$ and have
confirmed that the result does not change.  Therefore, in the rest of
this paper, we adopt the isothermal equation of state with
$T=10^4\cmr{K}$ in cases without radiation, to save computational costs.

After turning on the radiation, the accretion rate decreases further, as
seen in Fig.~\ref{fig:mdot_L_FB_fid}(a).  It reaches the smaller
constant value $\dot{M}= 5.0\times10^{-6}\,M_\odot\cmr{yr^{-1}}$ at
$t_\mr{end}=1.2\times10^6\cmr{yr}$, which is about 0.1 of the value before
turning on the radiation and even less than $0.01$ of the Bondi rate.
As shown in \cite{Sugimura:2017ab}, the accretion rate with this
parameter set but without rotation is very high with
$\dot{M}\sim0.6\,\dot{M}_\mr{B}$. Therefore, this result demonstrates that
the accretion rate is largely reduced by the interplay of angular
momentum and radiation.  The amount of the reduction will be
analytically understood in Sec.~\ref{sec:formulation}.
In Fig.~\ref{fig:mdot_L_FB_fid}(b), we see that the luminosity behaves
in the same way as $\dot{M}$ following Eq.~\eqref{eq:3}. The luminosity
$L$ is generally sub-Eddington, with $L\approx 0.2\,L_\mr{E}$ at
$t_\mr{end}$.

Fig.~\ref{fig:RSc01a1em2} shows the flow structure at $t_\mr{end}$.  The
accretion occurs through the neutral disc remaining inside the shadow.
The boundary between the ionized and neutral regions is determined by
the shadow angle, $\sim 45^\circ$ from the equatorial plane. Outside the
Bondi radius for the ionized gas, $r_\mr{B,HII}$, material on the
surface of the neutral gas is photoevaporated and flows out in the
vertical direction, as the sound speed, $c_\mr{s,HII}$, is larger than
the escape velocity, $v_\mr{esc}=(2GM_\mr{BH}/r)^{1/2}$.  Conversely,
inside $r_\mr{B,HII}$, where $c_\mr{s,HII}<v_\mr{esc}$, the photoionized
gas falls back to the disc again.  The outflows in the polar directions
blow away the low-$j$ gases initially locating near the poles.  Recall
that accretion of such material boosted $\dot{M}$ before the radiation
is turned on.

In Fig.~\ref{fig:RSc01a1em2}(c), a low-density region appears near the
poles as a result of the centrifugal barrier.\footnote{The temperature
is also low due to the inefficient photoionization heating, whose
heating rate is proportional to the recombination rate and hence the
square of the density.}  However, structures along the $z$-axis are
partly artifacts of assumed axisymmetry
\citep[e.g.,][]{Sugimura:2017ab}.  In addition, jets or outflows
launched near the BH, if included, would largely change such
features. Since these structures hardly affect $\dot{M}$ anyway because
of the small solid angle, we do not attempt to study them in more detail
below.

Although the ionization boundary looks similar in shape to that in the
non-rotating case, the flow structure is significantly altered by the
angular momentum.  Whereas the gas in the shadowed region falls freely
in the non-rotating case, it has to pass through the accretion disc
slowly in a viscous timescale otherwise. In the non-rotating case with
large $M_\mr{BH}$, \cite{Takeo:2018aa} found that, for a massive enough
($\sim 10^5\,M_\odot$) BH surrounded by a non-rotating medium, the
radiation is obscured by the gas which is pushed by the ram pressure and
intrudes into the polar regions, thereby mitigating the radiation
feedback. In the rotating case, however, the ram pressure of the flow is
so weak that we do not see such a phenomenon.  We have confirmed this by
performing an additional simulation with $M_\mr{BH}$ enhanced to
$10^5\,M_\odot$.

%%%%%%%%%%%%%%%%%%%%%%%%%%%%%%%%%%%%%%%%%%%%%%%%%%%%%
\subsection{Isotropic radiation run} \label{sec:rad_iso}
%%%%%%%%%%%%%%%%%%%%%%%%%%%%%%%%%%%%%%%%%%%%%%%%%%%%%

Next, we describe the isotropic radiation run. The set-up is the same as
in the previous run, except that the radiation is isotropic.

Figs.~\ref{fig:mdot_L_FB_fid}(a) and (b) show the time evolution of
$\dot{M}$ and $L$, respectively. Fig.~\ref{fig:mdot_L_FB_fid}(a) shows
that after the radiation is turned on, $\dot{M}$ oscillates violently
repeating burst and quiescent phases. This behaviour is similar to what
is found in the non-rotating case \citep[e.g.,][]{Sugimura:2017ab}, but
the averaged accretion rate in the last $10^5$ years, $\dot{M}=1.9\times
10^{-7}\,M_\odot\cmr{yr^{-1}}$, is about 10 times smaller.  Here, we
see again the reduction of $\dot{M}$ by the interplay of angular
momentum and radiation.
Again, Fig.~\ref{fig:mdot_L_FB_fid}(b) shows that $L$ oscillates in the
same way as $\dot{M}$ following Eq.~\eqref{eq:3}.  Its absolute value is
generally small and $L\ll L_\mr{E}$ even at the burst phases.

Figs.~\ref{fig:RIc01a1em2_BBAB} (a, b) and (c, d) show the gas
distribution before and after an accretion burst, respectively.  The
\Hii region contracts in a quiescent phase, and expands in a burst phase
\citep[e.g.,][]{Park:2011aa}. The neutral disc initially formed
(Fig.~\ref{fig:NRc01a1em2}) is completely ionized and vanishes soon
after the radiation is turned on.  The \Hii region is elongated along
the $z$-axis (Figs.~\ref{fig:RIc01a1em2_BBAB}a and c), because the
density decreases near the axis.

%%%%%%%%%%%%%%%%%%%%%%%%%%%%%%%%%%%%%%%%%%%%%%%%%%%%%
\subsection{Cases with different sets of the parameters}
\label{sec:iso_parm}
%%%%%%%%%%%%%%%%%%%%%%%%%%%%%%%%%%%%%%%%%%%%%%%%%%%%%

\begin{figure}
\centering \includegraphics[width=8cm]{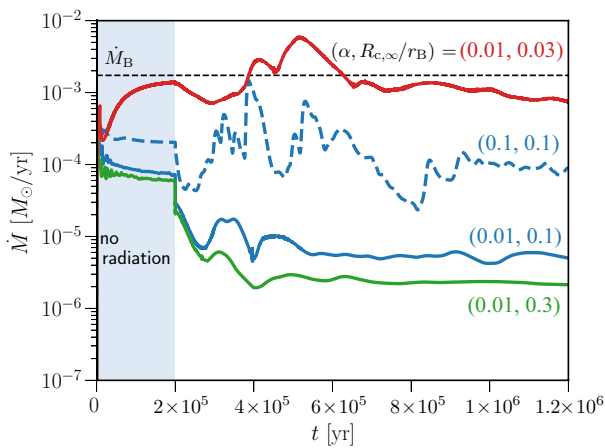}
\caption{Same as Fig.~\ref{fig:mdot_L_FB_fid}(a) but for the anisotropic radiation runs with
different parameter sets. The colours represent $R_\mr{c,\infty}/r_\mr{B}= 0.03$ (red), $0.1$ (blue) and $0.3$ (green), while the line types represent $\alpha=0.01$ (solid) and $0.1$ (dashed).
The run with $(\alpha,\ R_\mr{c,\infty}/r_\mr{B})=(0.01,\ 0.1)$ is also presented in Fig.~\ref{fig:mdot_L_FB_fid}(a).
}  \label{fig:mdot_FB_var}
\end{figure}

As described in Sec.~\ref{sec:cases}, we also study a number of cases
with different parameters of $(\alpha,\
R_\mr{c,\infty}/r_\mr{B})=(0.01,\ 0.03)$, $(0.1,\ 0.1)$ and $(0.01,\
0.3)$, in addition to the fiducial case presented above. We here only
study the evolution of the accretion rates for these cases. More
detailed analyses are provided later in Sec.~\ref{sec:formulation},
where we develop an analytical model that well explains the numerical
results for a wide range of the parameters.

Fig.~\ref{fig:mdot_FB_var} shows the time variation of $\dot{M}$
in the anisotropic radiation runs with the various parameter sets (see Table~\ref{tab:model}).  By the time radiation turns on
($t\leq2\times10^5\cmr{yr}$), the accretion flows reach steady sates. At
that time, $\dot{M}$ is larger with larger $\alpha$, while it is larger
with smaller $R_\mr{c,\infty}$ with weak dependence for
$R_\mr{c,\infty}/r_\mr{B}\gtrsim 0.1$. After the radiation turns on, the
flows reach different steady states by the end of simulations ($t=1.2\times 10^6\cmr{yr}$).  The final accretion rates depend on $\alpha$ and $R_\mr{c,\infty}$ in a similar way, but with the greater variation in this case.

In the isotropic radiation runs, $\dot{M}$ oscillates violently as
shown in Sec.~\ref{sec:rad_iso}, with the average accretion rates for
$3\times 10^5\cmr{yr}<t<4\times 10^5\cmr{yr}$ given in
Table~\ref{tab:model}.  They are very small in all three runs, with the
dependence on $\alpha$ and $R_\mr{c,\infty}$ similar to those obtained
above.  In this case, the dependence is as weak as before radiation
turns on.

%%%%%%%%%%%%%%%%%%%%%%%%%%%%%%%%%%%%%%%%%%%%%%%%%%%%%
%%%%%%%%%%%%%%%%%%%%%%%%%%%%%%%%%%%%%%%%%%%%%%%%%%%%%
\section{Analytical formulation}
\label{sec:formulation}
%%%%%%%%%%%%%%%%%%%%%%%%%%%%%%%%%%%%%%%%%%%%%%%%%%%%%
%%%%%%%%%%%%%%%%%%%%%%%%%%%%%%%%%%%%%%%%%%%%%%%%%%%%%

In the previous section, we found that the angular momentum can largely
suppress the accretion in the case with anisotropic radiation, where the
accretion rate would be high without angular momentum.  In order to
understand such suppression, we here develop an analytical model
describing accretion through a neutral disc connected to a medium.  This
model is motivated by the fact that the accretion occurs through the
neutral disc remaining inside the shadow in the anisotropic radiation
runs (Fig.~\ref{fig:RSc01a1em2}).  Note that this model is applicable to
neither the case with isotropic radiation where the disc is completely
photoionized (Fig.~\ref{fig:RIc01a1em2_BBAB}), nor that without
radiation where polar inflows of low angular momentum gas contribute to
the accretion (Fig.~\ref{fig:NRc01a1em2}). In the latter case, however,
such polar inflows might be prevented by jets or outflows launched near
the BH.

After developing the model, we derive the critical angular momentum of
the medium needed to suppress the accretion, as the model predicts that
the accretion rate goes back to the Bondi rate in the case with low
angular momentum.  Using the critical value, we can roughly estimate the
impact of the angular momentum on the accretion rate from the property
of the medium (e.g., Sec.~\ref{sec:growth}).

Below, we first develop the analytical model in Sec.~\ref{sec:analytic}.
Then, in Sec.~\ref{sec:understand}, we interpret the simulation results
obtained in Sec.~\ref{sec:results} with the model.  Furthermore, we
derive the condition for accretion suppression in Sec.~\ref{sec:cond}.

%%%%%%%%%%%%%%%%%%%%%%%%%%%%%%%%%%%%%%%%%%%%%%%%%%%%%
%%%%%%%%%%%%%%%%%%%%%%%%%%%%%%%%%%%%%%%%%%%%%%%%%%%%%
\subsection{Analytical model of  accretion 
through a disc connected to a medium} \label{sec:analytic}
%%%%%%%%%%%%%%%%%%%%%%%%%%%%%%%%%%%%%%%%%%%%%%%%%%%%%
%%%%%%%%%%%%%%%%%%%%%%%%%%%%%%%%%%%%%%%%%%%%%%%%%%%%%

Here, to understand the mechanism of the suppression of accretion by the
angular momentum, as well as to analytically estimate the extent of the
suppression, we develop an analytical model for the accretion through a
neutral disc connected to a rotating medium with constant specific
angular momentum $j=j_\infty$.  In this model, we suppose that the
accretion is suppressed because the gas stagnates at the centrifugal
radius and accumulates between the centrifugal and the Bondi radii, and
thus the pressure within the Bondi radius is enhanced.

We develop the model by connecting the following two types of solutions
at the centrifugal radius $R_\mr{c,\infty}$.  Outside $R_\mr{c,\infty}$,
where the dynamical equilibrium is held between the gravity, pressure
gradient and centrifugal force, the accretion is associated with the
inward gas supply that compensates the depletion to the inner disc at
$R_\mr{c,\infty}$.  Inside $R_\mr{c,\infty}$, where the Keplerian disc
is formed, the accretion is caused by the viscous angular momentum loss,
which transports the inward mass flux supplied at $R_\mr{c,\infty}$.
Below, we assume that the gas is isothermal.

We start by obtaining the surface density $\Sigma$ for the outer
dynamically equilibrium distribution that is connected to the medium
with $\rho=\rho_\infty$ \citep{Papaloizou:1984aa}.  Here, we assume that
the gas has $j=j_\infty$ everywhere. Then, the equation for the force
balance can be analytically solved, as shown in
Appendix~\ref{sec:stat_disc}.  Inside $r_\mr{B}$, where the scale height
$H_\mr{s}=c_\mr{s}/\Omega_\mr{K}$ is smaller than the radius $R$ and the
distribution is disc-like, $\Sigma$ is given by
(Appendix~\ref{sec:stat_disc})
\begin{align}
\Sigma = 
\frac{\sqrt{2\pi}c_\mr{s}}{\Omega_\mr{K}}\rho_\mr{\infty}
\exp\left[\frac{r_\mr{B}}{R} -\frac{R_\mr{c,\infty}r_\mr{B}}{2R^2}\right].
\label{eq:19}
\end{align}

Next, we obtain the relation between the accretion rate $\dot{M}$ and
$\Sigma$ for the inner viscous Keplerian disc
\citep[e.g.,][]{Shakura:1973aa,Kato:1998aa,Frank:2002aa}.  In
Appendix~\ref{sec:std_acc_disc}, assuming that the disc is thin and
adopting the $\alpha$-type viscosity, $\nu =
\alpha\,\gamma\,c_\mr{s}^2/\Omega_\mr{K}$, we obtain \citep[e.g., Eq.~5.19 of][]{Frank:2002aa}
\begin{align}
\dot{M} =3\pi \nu  \Sigma\,.
\label{eq:20}
\end{align}
In a steady accretion disc $\dot{M}$ becomes radially constant by
adjusting $\Sigma$.

We then connect the two solutions at $R_\mr{c,\infty}$ by substituting
Eq.~\eqref{eq:19} into Eq.~\eqref{eq:20}. We obtain
\begin{align}
 \dot{M}_\mr{suppr}&=
\frac{\sqrt{18\pi^3}\,\alpha\, \gamma\, c_\mr{s}\,R_\mr{c,\infty}^3}
{r_\mr{B}}\rho_\infty\exp\left[\frac{r_\mr{B}}{2R_\mr{c,\infty}}\right]\,,
\label{eq:28}
\end{align}
where we have used $\Omega_\mr{K}^2 %=GM_\mr{BH}/R_\mr{c,\infty}^3
=c_\mr{s}^2\,r_\mr{B}/R_\mr{c,\infty}^3$ at $R_\mr{c,\infty}$.  Here, if
$R_\mr{c,\infty}$ is so small that Eq.~\eqref{eq:28} yields
$\dot{M}_\mr{suppr}>\dot{M}_\mr{B}$, the assumption of the outer
dynamically-equilibrium distribution breaks down, because the gas cannot
be supplied to the disc at a rate exceeding $\dot{M}_\mr{B}$. Thus,
imposing that $\dot{M}$ does not exceed $\dot{M}_\mr{B}$, we modify
Eq.~\eqref{eq:28} as
\begin{align}
 \dot{M}=
\begin{cases}
 \dot{M}_\mr{suppr} &  (\dot{M}_\mr{suppr} <\dot{M}_\mr{B})\\
\dot{M}_\mr{B} & (\dot{M}_\mr{suppr} >\dot{M}_\mr{B})
\end{cases}\,,
\label{eq:26}
\end{align}
which gives an analytical estimate for the accretion rate from a
rotating medium.

\begin{figure}
\centering \includegraphics[width=8cm]{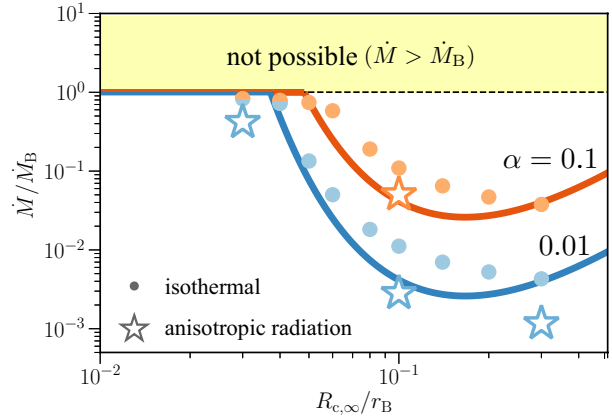}
\caption{The parameter dependence of $\dot{M}$ for the accretion from an
isothermal medium with $j=j_\infty$. We normalize $\dot{M}$ by $\dot{M}_\mr{B}$. 
The horizontal axis represents $R_\mr{c,\infty}\
(\equiv j_\infty^2/GM_\mr{BH})$, while the colors correspond to
$\alpha=0.1$ (orange) and 0.01 (blue). The lines are the analytical
estimates given by Eq.~\eqref{eq:26}, whereas the filled dots are the
numerical values obtained in Appendix~\ref{sec:iso_inidisc_sim}.  We
overplot the results of anisotropic radiation runs in
Sec.~\ref{sec:results} with open stars.}
\label{fig:mdot_iso}
\end{figure}

In Fig.~\ref{fig:mdot_iso}, the parameter dependence of the analytical
accretion rate $\dot{M}$ is shown, along with the numerical results
obtained in the equivalent settings (see
Appendix~\ref{sec:iso_inidisc_sim}). In the figure, we normalize
$\dot{M}$ by $\dot{M}_\mr{B}$.  Note that how much $\dot{M}$ is
suppressed with respect to $\dot{M}_\mr{B}$ depends only on the
combination of $\alpha$ and $R_\mr{c,\infty}/r_\mr{B}$ (not on either
$M_\mr{BH}$, $\rho_\infty$ or $c_\mr{s}$), because, with
Eqs.~\eqref{eq:6} and \eqref{eq:28}, $\dot{M_\mr{suppr}}/\dot{M}_\mr{B}$
can be rewritten as
\begin{align}
\frac{ \dot{M}_\mr{suppr}}{\dot{M}_\mr{B}}
=
2.8\,
\alpha \left(\frac{R_\mr{c,\infty}}{r_\mr{B}}\right)^3 \exp\left[\frac{r_\mr{B}}{2R_\mr{c,\infty}}\right]\,,
\label{eq:18}
\end{align}
where we have used $\gamma=5/3$.  The agreement of the analytical and
numerical results is remarkable considering the simplicity of the model.
In the case that the accretion is suppressed by the angular momentum,
i.e., $\dot{M}_\mr{suppr}<\dot{M}_\mr{B}$, the $\alpha$ dependence is
that $\dot{M}$ is proportional to $\alpha$ (Eq.~\ref{eq:28}).  As for
the $R_\mr{c,\infty}$ dependence, $\dot{M}$ rapidly increases with
decreasing $R_\mr{c,\infty}$ for $R_\mr{c,\infty}\lesssim 0.1\,
r_\mr{B}$, mainly because of the exponential increase of $\Sigma
(R_\mr{c,\infty})$ under the assumption of the dynamical equilibrium
(see Eq.~\ref{eq:19}).  When $\dot{M}_\mr{suppr}$ reaches
$\dot{M}_\mr{B}$, however, the dynamical equilibrium is broken and the
increase of $\dot{M}$ stops.  For lower $R_\mr{c,\infty}$, $\dot{M}$ is
equal to $\dot{M}_\mr{B}$, with little effect of the angular momentum.
The disagreement between the numerical and analytical results can be
partly attributed to the considerable thickness of disc at
$R_\mr{c,\infty}$ (see Fig.~\ref{fig:R_isofid}c), which is not
consistent with our assumption of thin inner disc. As the thickness at
$R_\mr{c,\infty}$ increases with $R_\mr{c,\infty}$ and the aspect ratio
reaches unity when $R_\mr{c,\infty}\sim r_\mr{B}$, our model is reliable
only when $R_\mr{c,\infty}$ is sufficiently smaller than $r_\mr{B}$.

%%%%%%%f%%%%%%%%%%%%%%%%%%%%%%%%%%%%%%%%%%%%%%%%%%%%%%
\subsection{Interpreting the simulation results with
the analytical model}
\label{sec:understand}
%%%%%%%%%%%%%%%%%%%%%%%%%%%%%%%%%%%%%%%%%%%%%%%%%%%%%

In this section, we understand the simulation results in
Sec.~\ref{sec:results} with the analytical model developed in the
previous section, focusing mainly on the case with anisotropic
radiation.  We begin with overplotting the simulation results (star
symbols) in Fig.~\ref{fig:mdot_iso}, where the parameter dependence of
$\dot{M}$ in the model (lines), as well as that for the isothermal
accretion (dot symbols), is shown. For all the four cases with
$(\alpha,\ R_\mr{c,\infty}/r_\mr{B})=(0.01,\ 0.1),\ (0.1,\ 0.01),\
(0.01,\ 0.03)\ \mr{and}\ (0.01,\ 0.3)$ examined in this paper, the
simulation results are well reproduced by the model.  Especially,
the transition between low and high-$\dot{M}$ regimes occurring at
$R_\mr{c,\infty}/r_\mr{B}\lesssim 0.1$ is clearly seen both in the
simulation results and the model. The accretion rate in the run with
$(\alpha,\ R_\mr{c,\infty}/r_\mr{B})=(0.01,\ 0.03)$ is $\sim
0.5\,\dot{M}_\mr{B}$, which is close to the value obtained with the
assumption of negligible angular momentum in the run with the same
anisotropic radiation field \citep{Sugimura:2017ab}. The downward
shifts of $\dot{M}$ by at most a factor of three compared to the
isothermal case (dot symbols) are partly attributable to the
photoevaporation mass-loss from the surface of the disc.  Note that we
have also confirmed that the equatorial gas profile in the anisotropic
radiation run with $\alpha=0.01$ and $R_\mr{c,\infty}=0.1\,r_\mr{B}$ is
consistent with that of the model (Appendix~\ref{sec:prof_aniso}).  The
agreement both in the accretion rate and the gas profile suggests that
the accretion is suppressed by the same mechanism as in the model, i.e.,
it is due to the stagnation of the gas at the centrifugal radius and the
consequent pressure enhancement within the Bondi radius.

Below, we make some remarks on the simulation results in the case without
radiation, as well as in the case with isotropic radiation (see Table~\ref{tab:model}).
In the former case, the accretion rates for the runs with
$(\alpha,\ R_\mr{c,\infty}/r_\mr{B})=(0.01,\ 0.1),\ (0.1,\ 0.1)\
\mr{and}\ (0.01,\ 0.3)$  are larger than in the model, because, in
addition to the accretion through the disc, the polar low-$j$ gas
directly flows into the sink and contributes to the accretion rate (see
Fig.~\ref{fig:NRc01a1em2}).  Recall that in the anisotropic radiation
runs, polar photoevaporative outflows prevent such inflows.  For
the run with $(\alpha,\ R_\mr{c,\infty}/r_\mr{B})=(0.01,\ 0.03)$,
however, $\dot{M}$ at $t=2\times10^5\cmr{yr}$ is
$0.8\,\dot{M}_\mr{B}$ and slightly lower than $\dot{M}_\mr{B}$ as
predicted by the model, because at that point of time $\dot{M}$ has yet
to reach the asymptotic value and is still on the rise.
Conversely, in the latter case, the accretion rates are smaller than in
the model, because the accretion disc is totally photoevaporated (see
Fig.~\ref{fig:RIc01a1em2_BBAB}).  In the anisotropic radiation runs,
however, the disc survives without being photoevaporated inside the
shadow.

%%%%%%%%%%%%%%%%%%%%%%%%%%%%%%%%%%%%%%%%%%%%%%%%%%%%%
\subsection{Condition for accretion suppression by the angular momentum}
\label{sec:cond}
%%%%%%%%%%%%%%%%%%%%%%%%%%%%%%%%%%%%%%%%%%%%%%%%%%%%%

\begin{figure}
\centering \includegraphics[width=8cm]{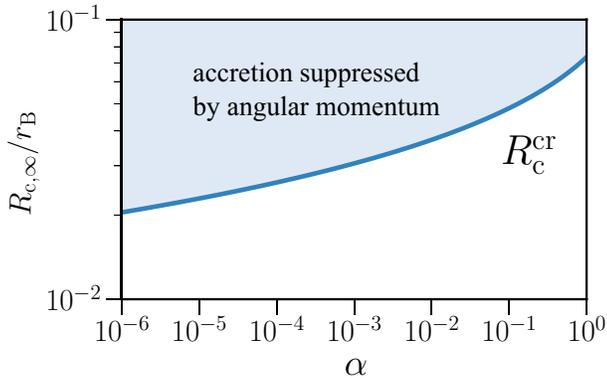} \caption{The
critical centrifugal radius $R_\mr{c}^\mr{cr}$ normalized by $r_\mr{B}$
as a function of $\alpha$.  The accretion is significantly suppressed by
the angular momentum in the shaded region ($R_\mr{c,\infty}>R_\mr{c}^\mr{cr}$).}
\label{fig:rc_cr}
\end{figure}

Next, we obtain the critical angular momentum above which the accretion
is significantly suppressed.

As mentioned in Sec.~\ref{sec:analytic}, we suppose that the effect of
angular momentum becomes negligible when Eq.~\eqref{eq:18} yields
$\dot{M}_\mr{suppr}/\dot{M}_\mr{B} > 1$.  Thus, we define the critical
centrifugal radius $R_\mr{c}^\mr{cr}$ by the condition
$\dot{M}_\mr{suppr}/\dot{M}_\mr{B}=1$. In Fig.~\ref{fig:rc_cr}, we plot
$R_\mr{c}^\mr{cr}$ as a function of $\alpha$.  The former slowly
increases with the latter, with $R_\mr{c}^\mr{cr}= 0.04\,r_\mr{B}$ at
$\alpha=0.01$ ($R_\mr{c}^\mr{cr}= 0.05\,r_\mr{B}$ at
$\alpha=0.1$). For a wide range of $\alpha$ with $10^{-6}<\alpha<1$, the
suppression of accretion becomes important as $R_\mr{c,\infty}$ exceeds
$O(10^{-2}\,r_\mr{B})$, with $\dot{M}$ decreased to
$O(\alpha\,\dot{M}_\mr{B})$ when $R_\mr{c,\infty}\sim 10^{-1}\,r_\mr{B}$
(see Fig.~\ref{fig:mdot_iso}). Recall that the case with
$R_\mr{c,\infty}\gtrsim r_\mr{B}$ is beyond the scope of our analytical
model (Sec.~\ref{sec:analytic}).

As expected with our analyses in Sec.~\ref{sec:understand},
Fig.~\ref{fig:rc_cr} agrees well with the numerical results
overall. Most of the anisotropic radiation runs examined are located in
the shaded region, showing the significant suppression of the accretion
rate. Only an exception is the case with
$(R_\mr{c,\infty},\alpha)=(0.03,0.01)$, which is really below the
critical line in Fig.~\ref{fig:rc_cr}.

%%%%%%%%%%%%%%%%%%%%%%%%%%%%%%%%%%%%%%%%%%%%%%%%%%%%%
\section{Discussion}
\label{sec:discussion}
%%%%%%%%%%%%%%%%%%%%%%%%%%%%%%%%%%%%%%%%%%%%%%%%%%%%%

%%%%%%%%%%%%%%%%%%%%%%%%%%%%%%%%%%%%%%%%%%%%%%%%%%%%%
\subsection{Growth of Pop III remnant BHs}
\label{sec:growth}
%%%%%%%%%%%%%%%%%%%%%%%%%%%%%%%%%%%%%%%%%%%%%%%%%%%%%

\begin{figure}
\centering \includegraphics[width=7cm]{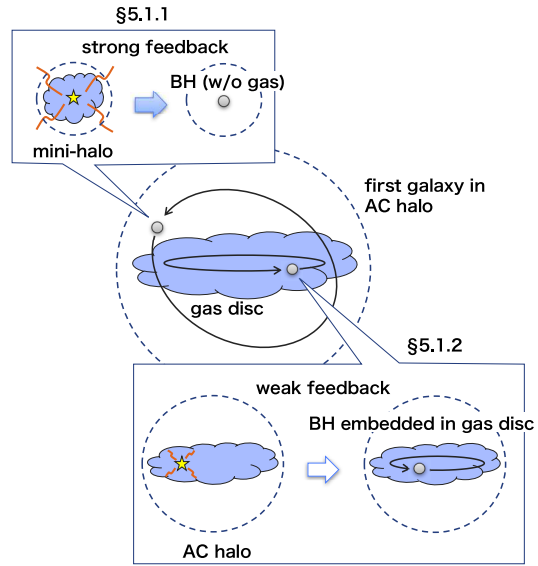}
\caption{Two evolutionary paths of a Pop III remnant BH and its
environment. A Pop III remnant BH can form either in a minihalo
(Sec.~\ref{sec:dis1}) or in an atomic-cooling (AC) halo
(Sec.~\ref{sec:dis2}). See text for details.} \label{fig:growth}
\end{figure}

Rapid growth of BHs is critically important for the formation of SMBHs
in the early Universe, especially if their growth starts from ``light
seeds'', i.e., the Pop III remnant BHs.  Reflecting the expected
diversity of the stellar mass of the Pop III stars
\citep[e.g.,][]{Hirano:2014aa,Hirano:2015aa}, their remnant BHs would
have a variety of masses, $\sim 100 - 10^3~M_\odot$.  The subsequent
growth of such BHs should depend on the environments surrounding them
(Fig.~\ref{fig:growth}).  In the standard bottom-up structure formation
paradigm in the $\Lambda$CDM Universe, Pop III stars normally form in
mini-halos with masses of $M_{\rm h}\simeq 10^{5-6}~M_\odot$ at $z>20$
\citep[e.g,][]{Yoshida:2006aa}. In addition, Pop III stars also form in
atomic-cooling halos with $M_{\rm h}\ga 3\times 10^7~M_\odot$, if prior
star formation is hampered by Lyman-Werner radiation and/or baryonic
streaming motions
\citep[e.g.,][]{Wise:2007ab,Visbal:2014ab,Tanaka:2014ab,Hirano:2017ac,Schauer:2017aa}.
In the following, we discuss the possibility of rapid growth of Pop III
remnant BHs, separately considering their different formation sites of
mini-halos and atomic-cooling halos.  We argue that the BH growth via
rapid mass accretion is not easily realized for either case.  In
particular, the angular momentum effect studied in the previous sections
is a critical obstacle for the growth of BHs formed in atomic-cooling
halos.

%-------------------------------------------------------------------------%

\subsubsection{Pop III remnants formed in mini-halos}
\label{sec:dis1}

Let us first consider the growth of Pop III remnant BHs in a mini-halo
with $M_{\rm h}\simeq 10^{5-6}~M_\odot$.  Massive Pop III stars with
$\ga 10^2-10^3~M_\odot$ produce intense ionizing radiation before
collapsing into BHs.  This ionizing radiation feedback evacuates a large
fraction of gas from the mini-halo, because of its shallow gravitational
potential\citep{Kitayama:2004aa}.  Rapid mass accretion on to the
remnant BH is thus not expected, at least when the BH is still harbored
in the same mini-halo \citep{Johnson:2007aa}.

%-----------------------------------------------------------------------%

Through the assembly of DM halos, a significant fraction of Pop III
remnant BHs fall into an atomic-cooling halo with $M_{\rm h} \ga
3\times10^7~M_\odot$ at $z\la 15$.  In such a massive and gas-rich halo,
the BHs could grow via accretion if they could sink to the galactic disc
due to dynamical friction.  We assume that a fraction $f_\star(\simeq
0.3)$ of the gas in the halo forms stars, i.e., $M_\star=f_{\star}
(\Omega_{\rm b}/\Omega_{\rm m})M_{\rm h}$, and those stars exert
friction on the remnant BHs.  Here, we adopt $\Omega_{\rm b}/\Omega_{\rm
m}=0.16$ \citep{Planck-Collaboration:2016aa}.  If the density
distribution of the stars is approximated by a singular isothermal
sphere, the dynamical friction timescale for a BH\footnote{Note that the
dynamical friction timescale could be shortened by a factor of two if the
friction is exerted by a gas instead of stars
\citep{Ostriker:1999aa,Escala:2004aa} } is estimated as
\citep{Binney:1987aa}

\begin{align}
t_{\rm DF}& \simeq \frac{1.17}{\ln \Lambda}~\frac{M_{\star}}{M_{\rm BH}} ~t_{\rm cross}, \nonumber\\
&\simeq 0.7~{\rm Gyr}~M_{\rm h,7.5}^{1/2} M_{\rm BH,3}^{-1} 
\left(\frac{f_\star}{0.3}\right)^{1/2} \left(\frac{r}{50~\mr{pc}}\right)^{3/2},
\end{align}
where the crossing time is $t_{\rm cross}=\sqrt{r^3/(GM_\star)}$ and the
Coulomb logarithm is set to $\ln \Lambda \sim \ln[M_\star/M_{\rm BH}]
\sim 6$.  In the above expression, we have defined $M_\mr{h,7.5}\equiv
M_\mr{h}/3\times10^7\,M_\odot$ and $M_\mr{BH,3}\equiv M_\mr{BH}/10^3\,M_\odot$.
We assume that the BH is located at the outskirts of the galactic disc,
i.e., $r\simeq  \lambda\, R_{\rm vir}$ \citep{Mo:1998aa},
where $\lambda \simeq 0.05$ is the dimensionless spin parameter and
$R_{\rm vir}\simeq 
1\cmr{kpc} ~M_{\rm h,7.5}(1+z)_{16}^{-1}$ is the
virial radius of the halo \citep{Barkana:2001aa}, with $(1+z)_{16} \equiv (1+z)/16$.  Using
this relation, the ratio of the dynamical friction timescale to the
Hubble timescale is estimated as
\begin{align}
\frac{t_{\rm DF}}{t_{\rm H}}& \simeq 
2~M_{\rm h,7.5}^{2} M_{\rm BH,3}^{-1} 
\lambda_{0.05}^{3/2} \left(\frac{f_\star}{0.3}\right)^{1/2},
\end{align}
with $\lambda_{0.05}=\lambda/0.05$. This suggests
 that dynamical friction is inefficient for BHs with 
$M_{\rm BH} \la 10^3~M_\odot$.
As a result, Pop III remnant BHs coming originally from mini-halos 
will just continue to wander in the outskirts of the galactic disc.
Such BHs will hardly grow via accretion, simply because the density of the surrounding gas is low ($n_\mr{H} \ll 10^4~{\rm cm}^{-3}$, also see \S\ref{sec:dis2} below).
In this case, the rapid growth of BHs is unlikely to occur regardless of whether the angular momentum of the gas affects the BH feeding.

%------------------------------------------------------------------------%

\subsubsection{Pop III remnants formed in atomic-cooling halos}
\label{sec:dis2}

Pop III stars can be also formed in a galactic disc in an atomic-cooling halo 
with $M_{\rm h}\ga 3\times10^7~M_\odot$ (or $T_{\rm vir} \ga 10^4~\mr{K}$).
Such a halo can hold the gas  against the stellar 
feedback \citep{Kitayama:2004aa}.
The remnant BHs are initially embedded in the gas-rich disc.

%------------------------------------------------------------------------%

Let us suppose that the BH is embedded in an isothermal, exponential disc
with a gas temperature $T_{\rm gas}\simeq 8000~\mr{K}$. 
The gas density at the mid-plane within the disc radius ($r\la \lambda R_{\rm vir}$) is estimated as \citep{Oh:2002aa}
\begin{align}
n_\mr{H} & \simeq 
1
\times 10^4~\mr{cm^{-3}}
~T_{\rm vir,4} \lambda_{0.05}^{-4}(1+z)_{16}^3
\left(\frac{f_{\rm d}}{0.3}\right)^2.
\end{align}
Assuming that a remnant BH has a peculiar velocity comparable to the circular velocity 
of the gas disc, 
\begin{align}
V & =\sqrt{\frac{G(M_{\rm d}+M_{\star})}{\lambda R_{\rm vir}}}\nnmb
&\simeq 
20
\,\mr{km\,s^{-1}}~T_{\rm vir,4}^{1/2} \lambda_{0.05}^{-1/2} \left(\frac{f_{\rm d}+f_\star}{0.6}\right)^{1/2},
\end{align}
which is higher than the sound speed of the gas 
$c_{\rm s}\simeq 
7
\,\mr{km\,s^{-1}}(T_{\rm gas}/8000~\mr{K})^{1/2}$.
Thus, the Bondi-Hoyle-Lyttleton (BHL) accretion rate is reduced by a factor of 
$\simeq [1+(V/c_{\rm s})^2]^{3/2}\simeq 
20$ from the value in Eq.~\eqref{eq:6}, 
\begin{align}
\dot{M}_{\rm BHL} & \simeq 
1
\times 10^{-5}~M_\odot\cmr{yr^{-1}}
\left(\frac{f_{\rm d}}{0.3}\right)^{1/2}
\left(\frac{f_{\rm d}+f_\star}{2f_{\rm d}}\right)^{-3/2}\nonumber\\
& ~~~ \times T_{\rm vir,4}^{-1/2}M_{\rm BH,3}^2
\lambda_{0.05}^{-5/2}
(1+z)_{16}^3.
\label{eq:BHL}
\end{align}
Note that Eq. (\ref{eq:BHL}) is valid for $(V/c_{\rm s})^2\gg 1$.  The typical BH
growth timescale is $t_{\rm grow}\sim M_{\rm BH}/\dot{M}_{\rm BHL}\sim
100~{\rm Myr}~M_{\rm BH,3}^{-1}(1+z)_{16}^{-3}$.  Thus, we obtain $t_{\rm
grow}/t_{\rm H}\sim 
0.3
~M_{\rm BH,3}^{-1}(1+z)_{16}^{-3/2}$, which
means that Pop III remnant BHs with $M_{\rm BH} \ga 3\times 10^2~M_\odot$ formed
inside an atomic-cooling halo may undergo rapid accretion.  {\it
However, the angular momentum of the accreting gas here comes into play
to prevent the BH growth.}

%--------------------------------------------------------------%

According to cosmological simulations of the first galaxy, 
the gas flow in an atomic-cooling halo is turbulent in general 
\citep[e.g.,][]{Wise:2007aa}.
We consider the same density fluctuation of the turbulent medium as in
our Galaxy, $\delta \rho/\rho \sim (L/2~\mr{pc})^{1/3}$, where $L$ is the 
characteristic spacial length of the fluctuation \citep{Armstrong:1995aa,Draine:2011aa}.
Accreting gas with such density fluctuation brings a net angular momentum, which
is estimated as
$\ell \sim (V r_{\rm B}/4)\cdot(\delta \rho/\rho)|_{L=2r_{\rm B}}$
\citep{Ipser:1977aa,Ioka:2017aa,Matsumoto:2018aa}.
Thus, the ratio of the centrifugal radius to the Bondi radius is given by
\begin{align}
\frac{R_{\rm c}}{r_{\rm B}}\simeq 
3
\times 10^{-2}~M_{\rm BH,3}^{2/3}
\left(\frac{V}{20\,\mr{km\,s^{-1}}}\right)^{-4/3}.
\end{align}
This ratio is comparable to the critical value $R_\mr{c}^{\rm cr}/r_{\rm
B}\sim 0.04$ obtained in Sec.~\ref{sec:cond}, above which the accretion
rate is suppressed by a factor of $\alpha \sim O(0.01\,\text{--}\,0.1)$,
but the former exceeds the latter as $M_\mr{BH}$ becomes larger than
$\sim 10^3\,M_\odot$. Thereafter, even if the BH is embedded in the gas
disc, its growth timescale becomes even longer than the age of the
Universe at $z=6$,
\begin{align}
t_{\rm grow}^{\rm ang}\sim
\frac{M_{\rm BH}}{\dot{M}_{\rm suppr}}\sim
10\cmr{Gyr}
~M_{\rm BH,3}^{-1}(1+z)_{16}^{-3}\alpha_{-2}\,,
\end{align}
where $\alpha_{-2}\equiv \alpha/10^{-2}$.
Therefore, it seems that {\it Pop III remnant BHs are hard to grow to
high-$z$ SMBHs via rapid accretion. }

%--------------------------------------------------------------------------------------%

\subsubsection{Possible pathways for rapid growth of Pop III remnant BHs}

A number of previous studies
\citep[e.g.,][]{Volonteri:2005aa,Madau:2014aa,Alexander:2014aa,Volonteri:2015aa,Tagawa:2015aa,Ryu:2016aa,Pacucci:2017ac}
have investigated the possibility of rapid (super-Eddington, i.e.,
$\dot{M} > 10\, \dot{M}_{\rm E}$ and thus $L > L_\mr{E}$) gas accretion
of Pop III remnant BHs to explain the existence of high-z SMBHs.  Since
our speculation seems different from those works, we discuss what
kind of uncertainties lead to the discrepancies, comparing
to our arguments in \S \ref{sec:dis1} and \ref{sec:dis2}.

\cite{Tagawa:2015aa} and \cite{Ryu:2016aa} considered the evolution of
Pop III remnant BHs in an atomic-cooing halo, performing N-body
simulations.  In fact, they concluded that some remnant BHs can fall
into the central region much faster than we estimated.  This is mainly
because the gas density is assumed to be higher than $\sim
10^4~\mr{cm^{-3}}$ or to follow an isothermal singular profile
($n_\mr{H} \propto r^{-2}$).  Because of the higher gas densities (at
smaller scales), dynamical friction allows the BHs to quickly sink into
the galactic center.  When the remnant BH reaches the central region with
$n_\mr{H}\gtrsim 10^6\cmr{cm^{-3}}\,M_{\rm BH,3}^{-1}$, the Bondi
accretion rate on to it becomes high enough ($\sim
5000~L_{\rm E}/c^2$) to realize hyper-Eddington accretion without
impeded by radiation feedback
\citep{Inayoshi:2016ac,Sakurai:2016aa,Sugimura:2017ab,Takeo:2018aa}.
This process may quickly form intermediate massive BHs with $\sim
10^5~M_\odot$ at the centre of the protogalaxy.

%-----------------------------------------------------------------%

Obviously, a key uncertainty is the density structure of the gas
containing the BHs.  \cite{Tagawa:2015aa} and \cite{Ryu:2016aa} assume
that the density profile does not change during the orbital evolution of
BHs.  In reality, however, the star formation will easily occur in the
high-density regions, so that the original density structure could be
modified by feedback effects such as supernova explosions
\citep[e.g.,][]{Dubois:2015aa,Yajima:2017ac}.  In future work, we
will study the evolution of the BH orbital motions and ambient
density structure, self-consistently incorporating the feedback effects.
Such treatment will also allow us to accurately estimate how much
angular momentum is brought by the accreting gas, and to assess whether
super-Eddington accretion is possible circumventing the angular momentum
barrier presented in this paper.

%%%%%%%%%%%%%%%%%%%%%%%%%%%%%%%%%%%%%%%%%%%%%%%%%%%%%
\subsection{Caveats}
\label{sec:caveats}
%%%%%%%%%%%%%%%%%%%%%%%%%%%%%%%%%%%%%%%%%%%%%%%%%%%%%

We have made a number of simplifications and approximations in this
work, and now discuss their significances.

First, we have examined only the two limiting cases of the anisotropy of
radiation field.  In the case of a non-rotating medium,
\cite{Sugimura:2017ab} have found that there is a critical shadowing angle
($\sim 10^\circ$ from the equatorial plane) above which the efficient
accretion is realized by the neutral Bondi-like inflows through the
equatorial layer that exceeds the photoevaporative mass-loss from the
surfaces. In the rotating case, however, the above condition needs to be
modified, because the photoevaporative mass-loss from the surfaces of
the rotationally-supported disc can be significant.  We have seen that
the disc is completely ionized by the isotropic radiation, while it is
not photoevaporated inside the $\sim45^\circ$ shadow in the case of the
anisotropic radiation. We will study the dependence of $\dot{M}$ on the
anisotropy more in the future.

Second, the actual anisotropy created inside the sink is highly
uncertain, although $\dot{M}$ strongly depends on it.  In the
literature, generation of (failed) winds or coronae above the disc have
been investigated
\citep[e.g.,][]{Hollenbach:1994aa,Begelman:1983aa,Woods:1996aa,Proga:2000aa,Wada:2012aa,Suzuki:2014aa,Nomura:2016aa}.
We expect that the materials associated with such structures obscure the
outward radiation and create its anisotropy.  In a future work, we plan to
study such process, considering its dependence on $M_\mr{BH}$,
$\dot{M}$, the metallicity of gas, etc..

Third, we have studied the idealized system of a static BH embedded
in a homogeneous medium, in order to understand how angular
momentum and radiation feedback affect the accretion flow. In considering more realistic BH accretion systems, however,
we need to take into account the effects of turbulence
\citep[e.g.,][]{Krumholz:2006aa,Hobbs:2011aa} and/or galactic-scale
inflows \citep[e.g.,][]{Hobbs:2012aa,Park:2016ab}.  In a highly
symmetric system as studied by our axisymmetric 2D simulations, the
angular momentum transported outward through a disc may accumulate near
the disc outer edge, resulting in the reduction of the accretion
rate. We in fact confirm this effect for the cases where the accreting
gas has small angular momentum ($R_\mr{c,\infty} \ll 0.1\,r_\mr{B}$; see
Appendix~\ref{sec:iso_inidisc_sim}).  In the case of accretion from a
turbulent medium, however, such accumulation may not occur because the
disc can change its rotational axis
before the accumulation proceeds, as the angular
momentum vector of the accreting gas varies in time.  Recall that in
this work, we have artificially removed the accumulated angular momentum
by imposing upper bound on specific angular momentum.  This procedure
may qualitatively mimic the above mechanism that works in a turbulent
medium.

Fourth, we have adopted the $\alpha$-type viscosity to mimic the
angular momentum transport via the turbulence driven by the
MRI. Although our results depend on the value of $\alpha$, as well as
where the viscosity works, i.e., the confinement factor $f$ in
Eq.~\eqref{eq:16}, it is computationally too expensive to perform 3D
magnetohydrodynamics simulations of the same problem. The unstable
non-axisymmetric modes can also affect the flow in the 3D simulations
\citep{Papaloizou:1984aa}. In the cases studied here, the Toomre $Q$
parameter is above unity and the disc is gravitationally stable.  In a
case with different parameter set, however, the gravitational
instability can play a role in transporting the angular momentum
depending on $M_\mr{BH}$ and $n_\mr{H,\infty}$ (see
Appendix~\ref{sec:toomreQ_iso}). It is also likely that the
Rayleigh-Taylor instability of the HII bubble, as seen in our 2D
simulations (e.g., Fig.~\ref{fig:RIc01a1em2_BBAB}), grows differently in
3D. The former 3D simulations \citep{Park:2017ab},
however, suggest that such difference does not significantly change the
accretion rate.  In addition, the disc is known to be Rayleigh-unstable
when the angular momentum decreases outward, so that the accumulated
angular momentum would be transported in 3D simulations \citep[see,
e.g.,][and reference therein]{Inayoshi:2018aa}.

Finally, although we assume that the dominant cooling process in the
neutral gas is the Ly$\alpha$ cooling, the $\mr{H}^{-}$ free-bound
cooling becomes dominant and cools the gas to $\sim 4\times10^3\cmr{K}$
when $n_\mr{H}\gg 10^6\cmr{cm^{-3}}$ \citep{Omukai:2001aa}. In
addition, the temperature might drop even to $\sim 2\times10^2\cmr{K}$
if $\mr{H}_2$ molecules somehow form in spite of the UV irradiation from
the BH neighbourhood.  Consideration of these processes could lead to
modification of $\dot{M}$.

%%%%%%%%%%%%%%%%%%%%%%%%%%%%%%%%%%%%%%%%%%%%%%%%%%%%%
\section{Summary}
\label{sec:conclusion}
%%%%%%%%%%%%%%%%%%%%%%%%%%%%%%%%%%%%%%%%%%%%%%%%%%%%%

We have investigated the combined effect of gas angular momentum and
radiation feedback on seed BH accretion, by preforming a suit of 2D
axisymmetric simulations considering both finite gas angular momentum
and radiation from the circum-BH disc.  The BH is located at the center
of a rotating medium, whose centrifugal radius is typically a tenth
of the Bondi radius.  We follow the formation of the
rotationally-supported disc, through which the accretion proceeds by the
angular momentum transport due to the assumed $\alpha$-type viscosity.

We have found that the accretion is strongly suppressed by the gas
angular momentum.  Except for the case with very low angular momentum, the
accretion rate is reduced by one order of
magnitude even without radiation feedback and becomes even smaller
with the feedback.  In particular, the accretion rate in the
case with anisotropic radiation field, which would be in the same order
as the Bondi rate without gas rotation \citep{Sugimura:2017ab}, is
reduced by a factor of $\alpha\sim O(0.01\,\text{--}\,0.1)$.  Our results
clearly indicate the importance of the interplay of the angular momentum
and radiation feedback.

We have also developed an analytical model that describes accretion
through a neutral disc connected to a medium. This model is capable of
reproducing the accretion rate obtained in the simulations with
anisotropic radiation. Furthermore, the model suggests the presence of the critical angular momentum above which the accretion is significantly suppressed.
The corresponding critical centrifugal radius normalized by the Bondi radius
is a weakly increasing function of $\alpha$
and equal to $0.04$ for $\alpha=0.01$,
suggesting that even such small angular momentum is enough to reduce the accretion rate.
This provides a useful estimate for the impact
of the angular momentum on the accretion rate, for example, in
future cosmological simulations of SMBH formation.

Finally, we have discussed the implications of our findings on the
growth of Pop III remnant BHs.  In the literature, those BHs are claimed
to grow to high-$z$ SMBHs by very rapid (super-Eddington)
accretion. However, the angular momentum effect studied in this
paper can be a crucial obstacle for such rapid mass accretion.
Whilst the condition for the formation of direct-collapse BHs is known
to be difficult to achieve \citep[e.g.,][]{Sugimura:2014aa}, it
should be equally challenging for the Pop III remnant BHs to rapidly
grow via the super-Eddington accretion.  Clearly, further studies are
needed to unveil the nature.

%%%%%%%%%%%%%%%%%%%%%%%%%%%%%%%%%%%%%%%%%%%%%%%%%%%%%%
\section*{Acknowledgements}
%%%%%%%%%%%%%%%%%%%%%%%%%%%%%%%%%%%%%%%%%%%%%%%%%%%%%%
The authors would like to thank Ken Ohsuga, Sanemichi Takahashi and
Kenji Toma for fruitful discussions.  The numerical simulations were
performed on the Cray XC30 at CfCA of the National Astronomical
Observatory of Japan, as well as on the computer cluster, {\tt Draco},
at Frontier Research Institute for Interdisciplinary Sciences of Tohoku
University and on the Cray XC40 at Yukawa Institute for Theoretical
Physics in Kyoto University.  This work is supported in part by
MEXT/JSPS KAKENHI Grant Number 15J03873 (KS), 25800102, 15H00776 and
16H05996 (TH), 17H04827 (HY) and 17H01102 and 17H06360 (KO) and by the
Simons Foundation through the Simons Society of Fellows (KI).

%\bibliography{$HOME/physics/mybib} %$

\appendix
%%%%%%%%%%%%%%%%%%%%%%%%%%%%%%%%%%%%%%%%%%%%%%%%%%%%%
\section{Inner and outer solutions of  analytical model}
\label{sec:derivation}
%%%%%%%%%%%%%%%%%%%%%%%%%%%%%%%%%%%%%%%%%%%%%%%%%%%%%
%overview
Our model consists of an outer dynamically equilibrium distribution
(Sec.~\ref{sec:stat_disc}) and an inner viscous Keplerian disc
(Sec.~\ref{sec:std_acc_disc}), which are connected at the centrifugal
radius. Below, we describe the outer and inner solutions in this order.

%%%%%%%%%%%%%%%%%%%%%%%%%%%%%%%%%%%%%%%%%%%%%%%%%%%%%
\subsection{Outer  dynamically equilibrium distribution}
\label{sec:stat_disc}
%%%%%%%%%%%%%%%%%%%%%%%%%%%%%%%%%%%%%%%%%%%%%%%%%%%%%

Here, we describe an outer dynamically equilibrium distribution
connected to a homogeneous medium \citep{Papaloizou:1984aa}.  We assume
that the gas is isothermal and that the angular momentum $j=j_\infty$
everywhere.

We start by describing the equation for the dynamical equilibrium between
the gravity, pressure gradient and centrifugal force,
\begin{align}
 -\frac{1}{\rho}\bm{\nabla} p - \left(\frac{GM_\mr{BH}}{r^2}\right)\bm{\hat{r}} + \left(\frac{j_\infty^2}{R^3}\right)\bm{\hat{R}}=0\,,
\label{eq:1}
\end{align}
where $\bm{\hat{r}}$ and $\bm{\hat{R}}$ are the unit vectors in the $r$
and $R$ directions, respectively.  Using the isothermal equation of state, $p =
c_\mr{s}^2 \rho$, Eq.~\eqref{eq:1} can be rewritten as
\begin{align}
 \bm{\nabla}\left(
c_\mr{s}^2\ln \rho - \frac{GM_\mr{BH}}{r} +\frac{j_\infty^2}{2R^2}
\right)=0\,.
\label{eq:2}
\end{align}
Then, imposing the outer boundary condition of constant density except
near the pole, i.e., $\rho\to\rho_\infty$ as $r\to\infty$ with finite
$\theta$, we obtain
\begin{align}
\rho &= \rho_\infty\exp\left[\frac{GM_\mr{BH}}{r c_\mr{s}^2} -\frac{j_\infty^2}{2R^2 c_\mr{s}^2}\right]\nnmb
&=\rho_\infty\exp\left[\frac{r_\mr{B}}{r} -\frac{R_\mr{c,\infty}r_\mr{B}}{2R^2}\right]\,,
\label{eq:4}
\end{align}
where we have used $r_\mr{B}=GM_\mr{BH}/c_\mr{s}^2$ and
$R_\mr{c,\infty}=j_\infty^2/GM_\mr{BH}$ in the second equality.  The
density $\rho$ rapidly decreases towards the pole, i.e., as $R\to0$,
due to the assumed constant angular momentum.

The density profile corresponds to that of a thin disc at $R \ll
r_\mr{B}$.  Near the equatorial plane, where $r_\mr{B}/r=
r_\mr{B}/(\sqrt{R^2+z^2}) \approx (r_\mr{B}/R)(1-z^2/2R^2)$, we can
approximate Eq.~\eqref{eq:4} as
\begin{align}
\rho \approx \rho_\mr{eq}(R)\exp\left[-\frac{z^2}{2H_\mr{s}^2}\right]\,,
\label{eq:12}
\end{align}
with the equatorial density,
\begin{align}
\rho_\mr{eq} =\rho_\infty\exp\left[\frac{r_\mr{B}}{R} -\frac{R_\mr{c,\infty}r_\mr{B}}{2R^2}\right]\,,
\label{eq:23}
\end{align}
and the scale height,
\begin{align}
 H_\mr{s}=\frac{c_\mr{s}}{\Omega_\mr{K}}\,.
\label{eq:13}
\end{align}
At $R \ll r_\mr{B}$, where the aspect ratio
$H_\mr{s}/R=(R/r_\mr{B})^{1/2}$ is small, the majority of gas is
confined to a layer with $|z|\lesssim O(H_\mr{s})$. In such layer, the
approximate form of Eq.~\eqref{eq:12} is valid and the integration in
the $z$-direction yields the expression for surface density, 
\begin{align}
\Sigma\approx \sqrt{2\pi} H_\mr{s}\rho_\mr{eq}(R) \,,
\label{eq:30}
\end{align}
which can be rewritten as Eq.~\eqref{eq:19} with Eq.~\eqref{eq:12}.
Note that $\rho_\mr{eq}$ has its maximum at $R=R_\mr{c,\infty}$, as the
pressure gradient is balanced with the gravity at $R>R_\mr{c,\infty}$
but with the centrifugal force at $R<R_\mr{c,\infty}$.

%%%%%%%%%%%%%%%%%%%%%%%%%%%%%%%%%%%%%%%%%%%%%%%%%%%%%
\subsection{Inner viscous Keplerian disc}
\label{sec:std_acc_disc}
%%%%%%%%%%%%%%%%%%%%%%%%%%%%%%%%%%%%%%%%%%%%%%%%%%%%%

Next, we describe a thin Keplerian disc extended towards the BH
\citep[e.g.,][]{Shakura:1973aa,Kato:1998aa,Frank:2002aa}.  Through the disc, the
steady accretion is driven by the angular momentum transport via the
$\alpha$-type viscosity.  Again, we assume the isothermal gas.

The gas distribution is governed by the following equations.  Since the
disc is vertically hydrostatic, the $z$-dependence of density is the
same as Eq.~\eqref{eq:12} and thus the relation between $\rho_\mr{eq}$
and $\Sigma$ is given by Eq.~\eqref{eq:30} (though $\rho_\mr{eq}$ is
different from Eq.~\ref{eq:23}).  Radially, the gravity is balanced with
the centrifugal force with angular velocity
\begin{align}
 \Omega = \Omega_\mr{K}\,.
\label{eq:5}
\end{align}
The vertically-integrated mass conservation equation 
can be written as 
\begin{align}
2\pi\,R\,v_\mr{R}\,\Sigma=-\dot{M}\,,
\label{eq:8}
\end{align}
with the accretion rate $\dot{M}\ (>0)$.
Finally, the vertically-integrated angular momentum conservation
equation is given by
\begin{align}
-\dot{M}R^2\Omega_\mr{K} + 3\pi R^2\nu\Sigma\Omega_\mr{K} = \dot{J}\,,
\label{eq:17}
\end{align}
with the net angular momentum flux $\dot{J}\,(=\mr{const})$.  Note that
we have used Eqs.~\eqref{eq:5} and \eqref{eq:8} to obtain
Eq.~\eqref{eq:17}.  Finally, we adopt the $\alpha$-type viscosity 
\begin{align}
 \nu = \frac{\alpha\,\gamma\,c_\mr{s}^2}{\Omega_\mr{K}}\,,
\label{eq:25}
\end{align}
which is the same as Eq.~\eqref{eq:16} with $f=1$.

The structure of the disc is uniquely determined once $c_\mr{s}$,
$\alpha$, $\dot{M}$ and $\dot{J}$ are given.  Here, $\dot{J}$ can be
determined from the inner boundary condition: we impose the torque-free
condition, i.e., the second viscous stress term in Eq.~\eqref{eq:17} is
set to zero, at the inner boundary where the first advection term, which
is proportional to $R^{1/2}$, is also small.\footnote{In the literature,
the torque-free boundary condition is often imposed at the innermost
stable circular orbit (ISCO), which is located at $6GM_\mr{BH}/c^2$ in
the case of Schwarzschild BHs.}  Thus, neglecting $\dot{J}$ in the
right-hand side of Eq.~\eqref{eq:17}, we finally obtain $\dot{M} = 3\pi
\nu \Sigma$ (Eq.~\ref{eq:20}).  For given $\dot{M}$, with Eq.~\eqref{eq:20}, as well as
Eqs.~\eqref{eq:30}, ~\eqref{eq:5}, \eqref{eq:8} and \eqref{eq:25}, we
can straightforwardly derive
\begin{align}
\rho_\mr{eq}=\frac{GM_\mr{BH}\dot{M}}{3\sqrt{2\pi^3}\alpha\gamma c_\mr{s}^3R^3}\propto R^{-3} \,,
\end{align}
\begin{align}
v_R=-\frac{3\alpha \gamma c_\mr{s}^2\sqrt{R} }{2\sqrt{GM}}\propto R^{1/2} \,,
\end{align}
and
\begin{align}
v_\phi=\sqrt{\frac{GM}{R}}\propto R^{-1/2}\,, 
\end{align}
which motivate the inner boundary conditions adopted in our simulations (see
Sec.~\ref{sec:num_setting}).

%%%%%%%%%%%%%%%%%%%%%%%%%%%%%%%%%%%%%%%%%%%%%%%%%%%%%
%%%%%%%%%%%%%%%%%%%%%%%%%%%%%%%%%%%%%%%%%%%%%%%%%%%%%
\section{Numerical confirmation of analytical model}
\label{sec:iso_inidisc_sim}
%%%%%%%%%%%%%%%%%%%%%%%%%%%%%%%%%%%%%%%%%%%%%%%%%%%%%
%%%%%%%%%%%%%%%%%%%%%%%%%%%%%%%%%%%%%%%%%%%%%%%%%%%%%

\begin{figure}
\centering \includegraphics[width=8cm]{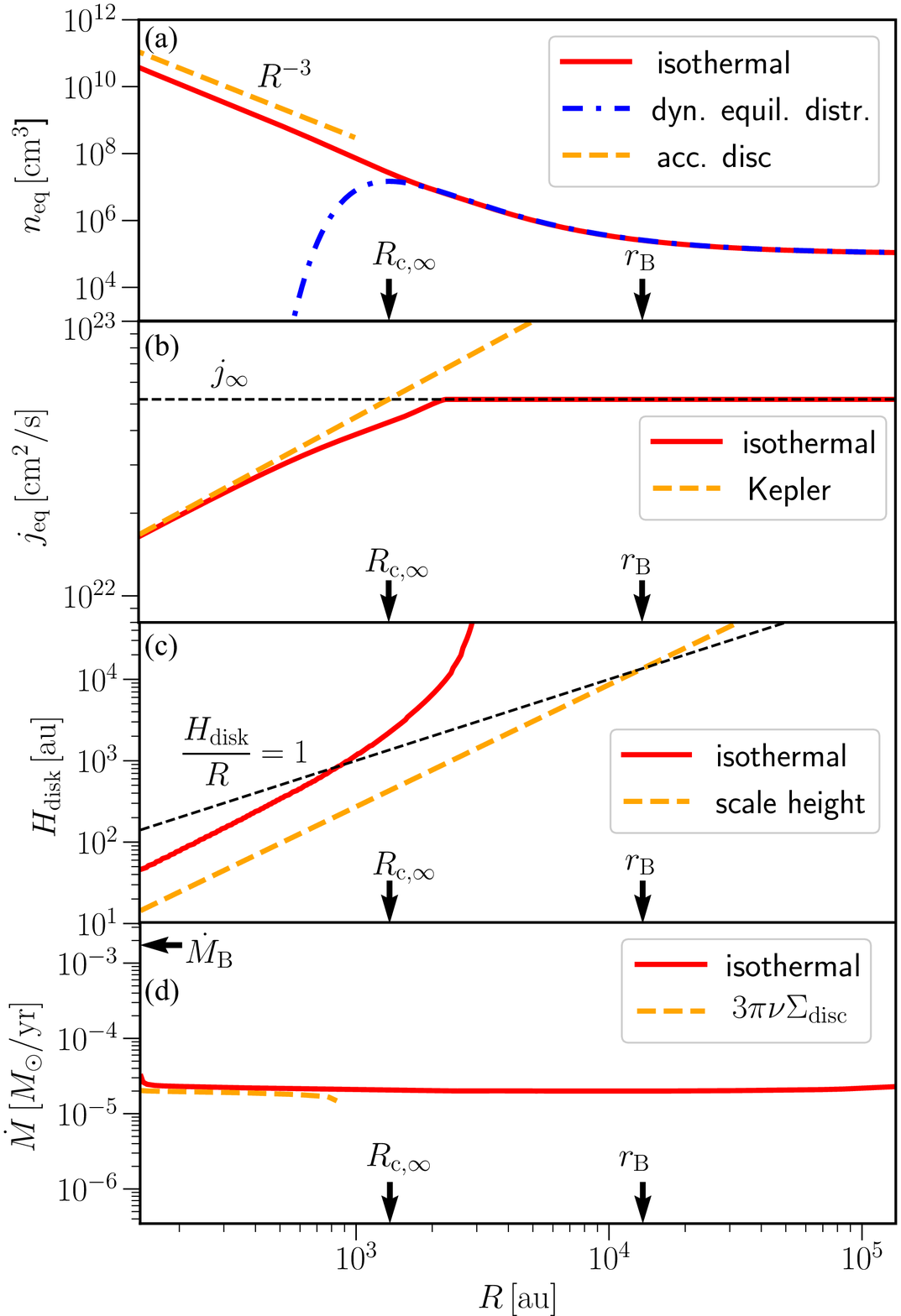}
\caption{The radial profiles of the gas in the isothermal accretion from
a medium with constant angular momentum.  Each panel represents (a)
equatorial density, (b) equatorial specific angular momentum, (c) disc
height and (d) net accretion rate. The case with $\alpha=0.01$ and
$R_\mr{c,\infty}=0.1\,r_\mr{B}$ is shown. } \label{fig:R_isofid}
\end{figure}

To confirm the validity of our analytical model, we compare the
analytical results with the numerical ones, by performing simulations in
the equivalent settings.

Here, we start the simulation from the dynamical equilibrium distribution
with $j=j_\infty$ everywhere (Eq.~\ref{eq:4}). For the computational
reason, if the density $\rho$ is initially below the floor density, we
set $\rho$ at the floor density and $j=0$.  We assume the isothermal gas
at $T=10^4\cmr{K}$.  Here, we adopt $\alpha=0.01$ or $0.1$ and
$R_\mr{c,\infty}/r_\mr{B}=0.03$, $0.04$, $0.05$, $0.06$, $0.08$, $0.14$,
$0.2$ or $0.3$.  We stop the calculation when the time variation of
$\dot{M}$ becomes sufficiently small. For the simulations presented
here, we limit $j$ to below $j_\infty$, as in the lowest angular momentum ($R_\mr{c,\infty}/r_\mr{B}=0.03$ ) run
in Sec.~\ref{sec:results}.
The rest of the numerical method is the same as that described in Sec.~\ref{sec:num_setting}.

To begin with, let us investigate the gas distribution, focusing on the
case with $\alpha=0.01$ and $R_\mr{c,\infty}=0.1\,r_\mr{B}$.  In
Fig.~\ref{fig:R_isofid}, the profiles of several quantities obtained at
the end of the simulation are compared with the analytical model.
First, Fig.~\ref{fig:R_isofid}(a) shows the equatorial density
$n_\mr{eq}$.  It agrees excellently with that of the dynamically
equilibrium distribution (Eq.~\ref{eq:23}) at $R>R_\mr{c,\infty}$, while
its $R$ dependence of $\propto R^{-3}$ is the same as the viscous
Keplerian disc at $R<R_\mr{c,\infty}$.
Second, in Fig.~\ref{fig:R_isofid}(b), the equatorial specific angular
momentum $j_\mr{eq}$ is presented.  It is equal to the asymptotic value
of $j_\infty$ at $R\gg R_\mr{c,\infty}$, while it is close to the
Keplerian profile $j_\mr{K}=R^2\Omega_\mr{K}$ at $R\ll R_\mr{c,\infty}$.
However, the agreement is not perfect around $R_\mr{c,\infty}$, because
the non-negligible pressure gradient reduces $j$ compared with
$j_\mr{K}$. Recall that we here impose the upper limit at $j_\infty$ by
removing $j$ otherwise accumulated just outside the outer edge of the
disc.
Third, in Fig.~\ref{fig:R_isofid}(c), we show the disc height $H_\mr{disc}$ at
which $\rho$ drops to one hundredth of $\rho_\mr{eq}$.  The disc is thin,
i.e., $H_\mr{disc}/R \lesssim 1$, at $R\lesssim R_\mr{c,\infty}$, but it
rapidly swells up at $R> R_\mr{c,\infty}$. For
the thin part, the relation $H_\mr{disc}\approx 3H_\mr{s}$ holds,
consistent with the vertical dependence of $\rho$ in Eq.~\eqref{eq:12}.
Finally, in Fig.~\ref{fig:R_isofid}(d), we plot the net accretion rate
over the entire solid angle,
\begin{align}
\dot{M} = 2 \int_0^{\pi/2} \rho(r,\theta) v_r(r,\theta) 2\pi r \sin \theta\,  \mr{d}\theta\,.
\label{eq:27}
\end{align}
It is constant with $R$, consistent with the steady accretion.  For the
range where the disc is thin ($H_\mr{disc}/R<1$), we also plot
$3\pi\nu\Sigma_\mr{disc}$ (see Eq.~\ref{eq:20}), using the disc surface
density, $\Sigma_\mr{disc}=\int_{-H_\mr{disc}}^{H_\mr{disc}} \rho(R,z)\,
\mr{d}z$, obtained from the simulation.  It agrees surprisingly well
with $\dot{M}$, implying that the accretion is indeed driven by the
angular momentum loss.  In summary, the numerical result is fully
consistent with the analytical model.

As expected from the above agreement in the gas distribution, the
analytical model can nicely reproduce the numerical accretion rates for
the various sets of the parameters, as shown in
Fig.~\ref{fig:mdot_iso}. In all cases, the difference between the
analytical and numerical results is less than a factor of three,
ensuring the qualitative validity of the analytical model.

In order to check the effect of imposing the upper limit on $j$, we have
made test calculations without the limit. In such a case, we have found
that $j$ is accumulated outside the outer edge of the disc, which
gradually expands as the accumulation proceeds.  The effective
enhancement of $j$ makes $\dot{M}$ smaller, especially in the case with
smaller $R_\mr{c,\infty}$. As a result, the rapid increase of $\dot{M}$ with
decreasing $R_\mr{c,\infty}$, as seen in Fig.~\ref{fig:mdot_iso},
disappears. The effect of imposing the limit is
insignificant for the case with $R_\mr{c,\infty} \gtrsim 0.1\,
r_\mr{B}$, where its impact on $\dot{M}$ is at most $30$ percent.

Finally, we briefly mention the effect of changing $\tilde\Omega$, which
regulates the region where the viscosity works (Eq.~\ref{eq:16}).  In
the test runs with $\tilde\Omega$ different from the fiducial value of
$0.8$ (with $\alpha=0.01$ and $R_\mr{c,\infty}=0.1\,r_\mr{B}$), 
we have seen that $\dot{M}$ increases with decreasing
$\tilde\Omega$ and becomes three times larger in the case of
$\tilde\Omega=0.6$. While the quantitative determination of $\dot{M}$ is
affected by the specific prescription of viscosity, we do not expect it
qualitatively changes the conclusion of this work.

%%%%%%%%%%%%%%%%%%%%%%%%%%%%%%%%%%%%%%%%%%%%%%%%%%%%%
%%%%%%%%%%%%%%%%%%%%%%%%%%%%%%%%%%%%%%%%%%%%%%%%%%%%%
\section{Equatorial  gas profile in anisotropic radiation run}
\label{sec:prof_aniso}
%%%%%%%%%%%%%%%%%%%%%%%%%%%%%%%%%%%%%%%%%%%%%%%%%%%%%
%%%%%%%%%%%%%%%%%%%%%%%%%%%%%%%%%%%%%%%%%%%%%%%%%%%%%

\begin{figure}
\centering \includegraphics[width=8cm]{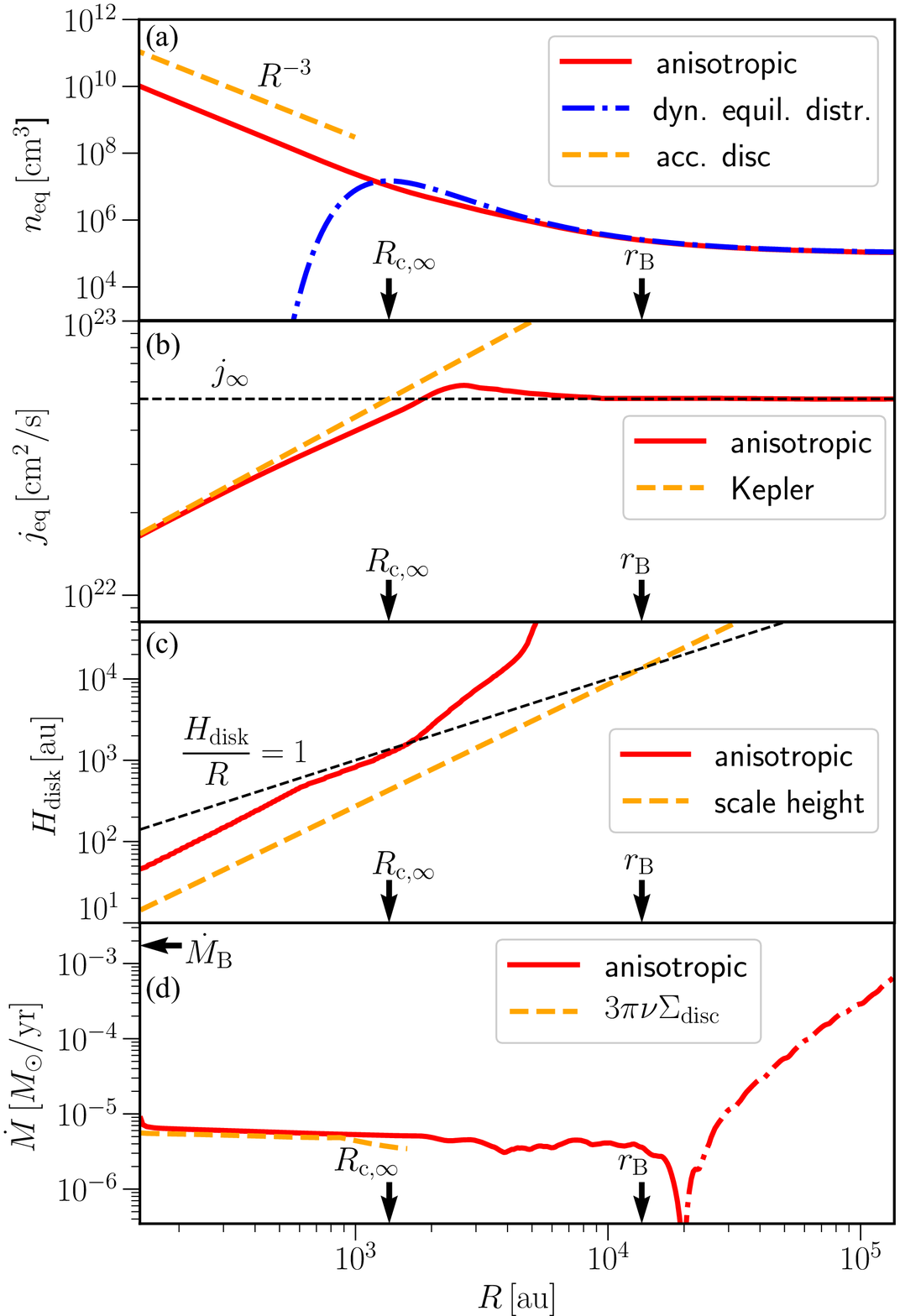} \caption{Same
as Fig.~\ref{fig:R_isofid} but for the anisotropic radiation run with
$\alpha=0.01$ and $R_\mr{c,\infty}=0.1\,r_\mr{B}$. In panel (d), solid (dot-dashed)
lines show a positive (negative) value.  }
\label{fig:RS_prof}
\end{figure}

Here, we show the equatorial gas profile in the anisotropic radiation
run in Sec.~\ref{sec:fiducial}, where $\alpha=0.01$ and
$R_\mr{c,\infty}=0.1\,r_\mr{B}$.  We average the profiles of
$n_\mr{eq}$, $j_\mr{eq}$, $H_\mr{disc}$ and $\dot{M}$ for the last
$3\times10^5\cmr{yr}$ with the time intervals of $2\times10^3\cmr{yr}$
(using $150$ snapshots in total) and plot them in Fig.~\ref{fig:RS_prof} in the
same way as Fig.~\ref{fig:R_isofid}.  In each panel, the profile in the
anisotropic radiation run generally agrees with that considered in the
analytical model.

Let us make several remarks on Fig.~\ref{fig:RS_prof}.  
First, in Fig.~\ref{fig:RS_prof}(c), while $H_\mr{disc}$ is smaller than
the assumed $45^\circ$ shadow and ionizing photons do not reach the disc
surfaces at $R\lesssim 0.5\,R_\mr{c,\infty}$, $H_\mr{disc}/R\sim 1$
at $R\sim R_\mr{c,\infty}$ because $H_\mr{disc}$ is determined by the
shadow angle there.
Second, in Fig.~\ref{fig:RS_prof}(b), $j_\mr{eq}$ exceeds $j_\infty$ at
$R\gtrsim R_\mr{c,\infty}$ due to the accumulation of $j$ transported
from the inside.  Note that, for simplicity, we do not impose the upper
limit on $j$ in this run.  Recall, however, that we have seen in
Appendix~\ref{sec:iso_inidisc_sim} that the imposition of the upper limit
does not significantly affect $\dot{M}$ for the case with 
$R_\mr{c,\infty}\gtrsim 0.1\,r_\mr{B}$.
Finally, in Fig.~\ref{fig:RS_prof}(d), the net accretion rate $\dot{M}$
is roughly constant at $R \lesssim r_\mr{B}$ but largely varies outside,
corresponding to the small time variation remaining even near the end
the simulation, as seen in Fig.~\ref{fig:mdot_L_FB_fid}(a).

%%%%%%%%%%%%%%%%%%%%%%%%%%%%%%%%%%%%%%%%%%%%%%%%%%%%%
\section{Dependence on numerical configurations}
\label{sec:resdep_iso}
%%%%%%%%%%%%%%%%%%%%%%%%%%%%%%%%%%%%%%%%%%%%%%%%%%%%%

\begin{figure}
\centering \includegraphics[width=8cm]{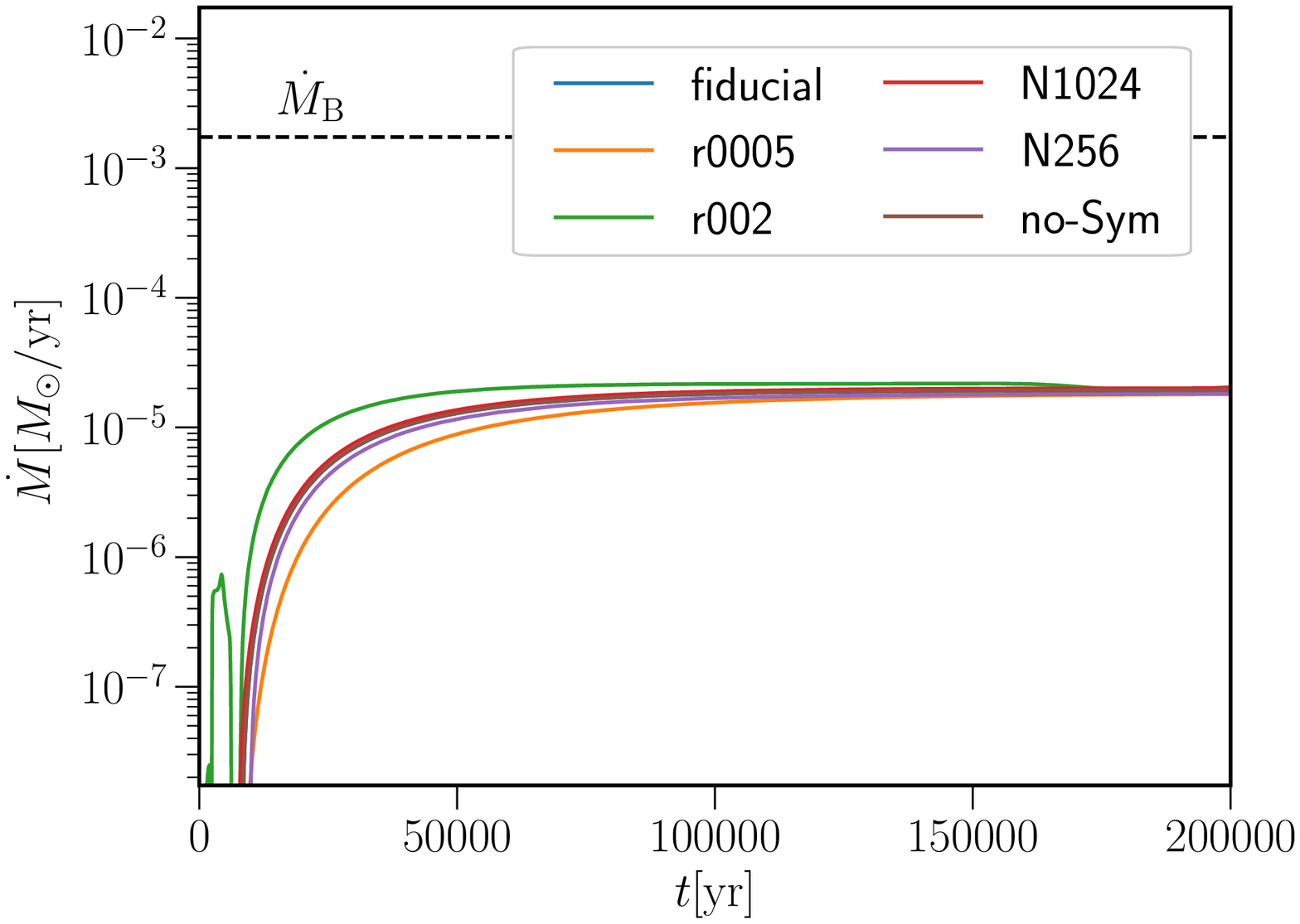}
\caption{Same as Fig.~\ref{fig:mdot_L_FB_fid}(a) but for the fiducial
case in Sec.~\ref{sec:iso_inidisc_sim} with different numerical
configurations (see text). The lines for the fiducial, no-Sym
and N1024 runs are overlapped with each other. } \label{fig:iso_numdep}
\end{figure}

To examine the dependence of our results on the numerical configuration,
we rerun the simulation of the fiducial case in
Appendix~\ref{sec:iso_inidisc_sim} with different numerical
configurations.  Until now, we adopt the following configurations: the
inner boundary is at $r_\mr{in}= 10^{-2}\,r_\mr{B}$; the number of grids
is $N_r\times N_\theta = 512\times 144$; and the range of $\theta$ is
$0<\theta<\pi/2$ under mid-plane symmetry. However, we here replace one
of them as: $r_\mr{in}=5\times 10^{-3}\,r_\mr{B}$ (r0005 run);
$r_\mr{in}=2\times 10^{-2}\,r_\mr{B}$ (r002); $N_r\times N_\theta =
1024\times 288$ (N1024); $N_r\times N_\theta =256\times 72$ (N256); and
$0<\theta<\pi$ without mid-plane symmetry (no-Sym).

Fig.~\ref{fig:iso_numdep} shows the time evolution of $\dot{M}$.  While
there are some differences in the early stage ($t\lesssim
5\times10^4\cmr{yr}$), all the runs give practically the same value in
the end.  Note that $\dot{M}$ is not affected by the assumption of
mid-plane symmetry, while the non-symmetric modes are known to be
evident in the former simulations of the accretion to active galactic
nuclei \citep[e.g.,][]{Stone:1999aa}. Note also that we find
considerable deviations between the cell-centered and cell-boundary mass
fluxes in the run with the reduced grids (N256), although $\dot{M}$ is
not affected by such deviations. In summary, our simulation results of
$\dot{M}$ are apparently independent of the numerical configuration in
the case without radiation feedback.

In this work, we consider the radiation feedback, as well as the angular
momentum. For the non-rotating case with radiation feedback,
\cite{Sugimura:2017ab} have shown that the resolution dependence is
insignificant with the current numerical configuration.  Thus, we expect
that the dependence of our results on the numerical configuration in the
case with both angular momentum and radiation feedback is also
modest.

%%%%%%%%%%%%%%%%%%%%%%%%%%%%%%%%%%%%%%%%%%%%%%%%%%%%%
\section{Gravitational stability of discs}
\label{sec:toomreQ_iso}
%%%%%%%%%%%%%%%%%%%%%%%%%%%%%%%%%%%%%%%%%%%%%%%%%%%%%

In the cases studied in this work, the disc is always gravitationally
stable according to the Toomre criterion, i.e., the Toomre $Q$ defined
as
\begin{align}
 Q = \frac{\kappa_\Omega c_\mr{s}}{\pi G \Sigma}\,,
\label{eq:9}
\end{align}
with the epicyclic frequency
$\kappa_\Omega=((2\Omega/R)\mr{d}/\mr{d}R(R^2\Omega))^{1/2}$, is larger
than unity. In the following, to see under what conditions the disc
becomes gravitationally unstable, we derive the parameter dependence of
$Q$ based on the analytical model in Sec.~\ref{sec:analytic}.

Using $\Sigma$ of the analytical model obtained from Eqs.~\eqref{eq:20}
and \eqref{eq:26}, we can rewrite Eq.~\eqref{eq:9} as
\begin{align}
 Q &\approx \frac{3 c_\mr{s}^2r_\mr{B}}
{2\pi^{\frac{3}{2}}G\,\,R_\mr{c,\infty}^3\,\rho_\infty}
\exp\left[-\frac{r_\mr{B}}{2R_\mr{c,\infty}}\right]\,,
\label{eq:21}
\end{align}
where the $R$ and $\alpha$ dependences are canceled out.  This yields
$Q\sim 10^3$ when $M_\mr{BH}=10^3\,M_\odot$,
$n_\mr{H,\infty}=10^5\cmr{cm^{-3}}$ and $R_\mr{c,\infty}=
0.1\,r_\mr{B}$, consistent with the simulation results.  With the Jeans
length $\lambda_\mr{J} = \sqrt{\pi c_\mr{s}^2/G\rho}$, Eq.~\eqref{eq:21}
can be further rewritten as $Q\approx 0.3\,(r_\mr{B}/R_\mr{c,\infty})^3
(\lambda_\mr{J}/r_\mr{B})^2\exp[-r_\mr{B}/2R_\mr{c,\infty}]$. This
suggests that $Q$ could be less than unity, i.e., the disc could be
gravitationally unstable, in the case with smaller
$\lambda_\mr{J}/r_\mr{B}\ (\propto M_\mr{BH}^{-1}
n_\mr{H,\infty}^{-1/2})$ and/or $R_\mr{c,\infty}/r_\mr{B}$.

\end{document}